\documentclass[runningheads]{llncs}
\usepackage{graphicx}
\usepackage{subfig}

\usepackage{color}
\definecolor{blue(pigment)}{rgb}{0.2, 0.2, 0.6}

\usepackage{lineno,hyperref}
\modulolinenumbers[5]

\usepackage{hyperref}
\hypersetup{
    colorlinks=true,
    citecolor=blue,
    linkcolor=blue,
    filecolor=magenta,
    urlcolor=blue
}
\usepackage{colortbl}%
  
\usepackage{xspace}
\usepackage{amsmath}
\usepackage{amssymb}
\usepackage{pifont}
\usepackage{footnote}
\usepackage{multicol}
\usepackage{multirow}
\usepackage{booktabs}
\usepackage{array,makecell}
\usepackage[export]{adjustbox}
\usepackage{arydshln}
\usepackage{graphicx}
\usepackage{subfig}
\usepackage{rotating}
\usepackage{wasysym}
\usepackage{soul}
\usepackage{xcolor}
\usepackage{color}
\usepackage{wrapfig}
\usepackage{caption}

\usepackage{pgfplots}
\pgfplotsset{width=10cm,compat=1.14}
\usepackage{tikz,pgfplotstable}
\usepackage[utf8]{inputenc}
\usepackage{pgfplots,pgfplotstable,booktabs}
\usetikzlibrary{positioning}

\definecolor{blue(pigment)}{rgb}{0.2, 0.2, 0.6}

\newcommand{\partitle}[1]{\vspace{2mm}\noindent\textbf{#1}}

\newcommand{\dataset}{Book-Crossing\xspace}

\newcommand{\ran}{\texttt{Random}\xspace}
\newcommand{\topp}{\texttt{MostPop}\xspace}

\newcommand{\uknn}{\texttt{UserKNN}\xspace}

\newcommand{\mf}{\texttt{MF}\xspace}
\newcommand{\pmf}{\texttt{PMF}\xspace}
\newcommand{\nmf}{\texttt{NMF}\xspace}
\newcommand{\wmf}{\texttt{WMF}\xspace}

\newcommand{\bpr}{\texttt{BPR}\xspace}
\newcommand{\pf}{\texttt{PF}\xspace}

\newcommand{\vaecf}{\texttt{VAECF}\xspace}
\newcommand{\neumf}{\texttt{NeuMF}\xspace}

\newcommand{\eg}{e.g., }
\newcommand{\ie}{i.e., }

\begin{document}
\title{The Unfairness of Popularity Bias in Book Recommendation}
\titlerunning{The Unfairness of Popularity Bias in Book Recommendation}
%

\author{Mohammadmehdi Naghiaei\inst{1} \and
Hossein A.~Rahmani\inst{2}\thanks{Corresponding author.} \and
Mahdi Dehghan\inst{3}}

\authorrunning{M.~Naghiaei, H.~A.~Rahmani, M.~Dehghan}

\institute{University of Southern California, USA \\ \email{naghiaei@usc.edu} \and
University College London, United Kingdom \\
\email{h.rahmani@ucl.ac.uk} \and
Shahid Beheshti University, Iran \\
\email{mahdi.dehghan551@gmail.com}}
\maketitle              
\begin{abstract}
Recent studies have shown that recommendation systems commonly suffer from popularity bias. Popularity bias refers to the problem that popular items (\ie frequently rated items) are recommended frequently while less popular items are recommended rarely or not at all. 
Researchers adopted two approaches to examining popularity bias: (i) from the users' perspective, by analyzing how far a recommendation system deviates from user's expectations in receiving popular items, and (ii) by analyzing the amount of exposure that long-tail items receive, measured by overall catalog coverage and novelty. In this paper, we examine the first point of view in the book domain, although the findings may be applied to other domains as well. To this end, we analyze the well-known \textit{Book-Crossing} dataset and define three user groups based on their tendency towards popular items (i.e., Niche, Diverse, Bestseller-focused). Further, we evaluate the performance of nine state-of-the-art recommendation algorithms and two baselines (\ie Random, MostPop) from both the accuracy (\eg NDCG, Precision, Recall) and popularity bias perspectives.
Our results indicate that most state-of-the-art recommendation algorithms suffer from popularity bias in the book domain, and fail to meet users' expectations with Niche and Diverse tastes despite having a larger profile size. Conversely, Bestseller-focused users are more likely to receive high-quality recommendations, both in terms of fairness and personalization. Furthermore, our study shows a tradeoff between personalization and unfairness of popularity bias in recommendation algorithms for users belonging to the Diverse and Bestseller groups, that is, algorithms with high capability of personalization suffer from the unfairness of popularity bias. Finally, across the models, our results show that \wmf and \vaecf can provide a higher quality recommendation when considering both accuracy and fairness perspectives.


\end{abstract}

\section{Introduction}
\label{sec:intro}
Recommender systems have been utilized in various information spaces such as entertainment, education, and online dating. They aim to support users in finding desired information that is typically hard and time-consuming to find without such systems. Recommendation algorithms proposed in the literature can be categorized into multiple groups according to the context of recommendation and how they learn user preference.

Collaborative Filtering (CF) is one of the most widely-investigated classes of algorithms used for recommendation. 
These algorithms generate recommendations based on explicit or implicit interactions between users and items in the system \cite{zhang2019deep,rahmani2020joint} without benefiting from the content information, as opposed to content-based recommendation techniques.
Hence, one limitation of CF algorithms is the problem of popularity bias which causes the popular (\ie short-head) items to be over-emphasized in the recommendation list. In contrast, the majority of less popular (\ie long-tail) items do not get enough visibility in the recommendation lists. Tackling popularity bias can make the recommender system more applicable in the real world for various reasons \cite{celma2008hits,ciampaglia2018algorithmic}.
A recommender system suffering from popularity bias would result in the market being dominated by a few well-known brands and deprive the discovery of new and unpopular items, which could ignore the interest of users with niche tastes.

Popularity bias has been largely investigated from the item-centered perspective, that is, how frequently popular items appear in recommendation lists. The item-centered perspective study ignores users' interest in popular or less popular items, which causes a limitation as the popularity bias does not affect users equally \cite{abdollahpouri2021user}. Recently, Abdollahpouri et al.~\cite{abdollahpouri2019unfairness} have conducted a research study to look at the popularity bias from a different perspective: the users'. To shed light on this topic, suppose a user rated $40$ long-tail (less-popular) items and $60$ popular items. Therefore, we expect the recommendation algorithm to end up with the same ratio of popular items in the recommended list presented to this user. Despite our expectations, most recommendation algorithms generate a recommendation list including close to $100$ popular items. Furthermore, Abdollahpouri et al.~\cite{abdollahpouri2019unfairness} evaluate how popularity bias leads to unfair treatment of different user groups according to their interests in popular movies. Their experiments demonstrate that the state-of-the-art recommendation algorithms can not comprehensively understand users who tend to be interested in unpopular items~\cite{abdollahpouri2019managing}. 

Popularity biases are known to arise from the underlying characteristics of the data, such as the number of interactions per item or user, sparsity, and an unbalanced distribution of interactions among items. As a result, in this work, we aim to reproduce the research study of Abdollahpouri et al.~\cite{abdollahpouri2019unfairness} and conduct it in a different domain and dataset. In particular, we select the book domain and \dataset dataset due to a variety of reasons. First, the \dataset data characteristic differs significantly from the previous study on the MovieLens 1M dataset in that it reports ($256.46$, $165.59$) interactions per item and users respectively, as opposed to ($12.79$, $13.92$) in \dataset dataset. Table \ref{tbl:dataset_char} summarizes the main data characteristic of \dataset dataset.
Additionally, there are several aspects of book recommendations that make them different, challenging, and somehow more important than recommendations entertainment industry, such as movie and music recommendations \cite{alharthi2018survey,deldjoo2021explaining}. For instance, although book-reading must become more prevalent in society, the investigation proves the opposite \ie the practice of reading for pleasure or education is declining, particularly among the young \cite{alharthi2018survey}. Also, since a reader spends much time reading books, it is crucial that the content matches their known preferences and expectations. The experience of finding a match that is suitable can be rewarding for readers, but an inappropriate recommendation could result in losing interest, further contributing to the downward trend. It is, therefore, crucial to examine the existence of popularity bias from a user-centered perspective in book recommendations and, for reasons of comparability, we would like to answer the same two research questions as in the study of Abdollahpouri et al.~\cite{abdollahpouri2019unfairness}:

\begin{figure}[t]
    \centering
    \subfloat[The long-tail of item popularity]
        {
            {
                \includegraphics[width=5cm]{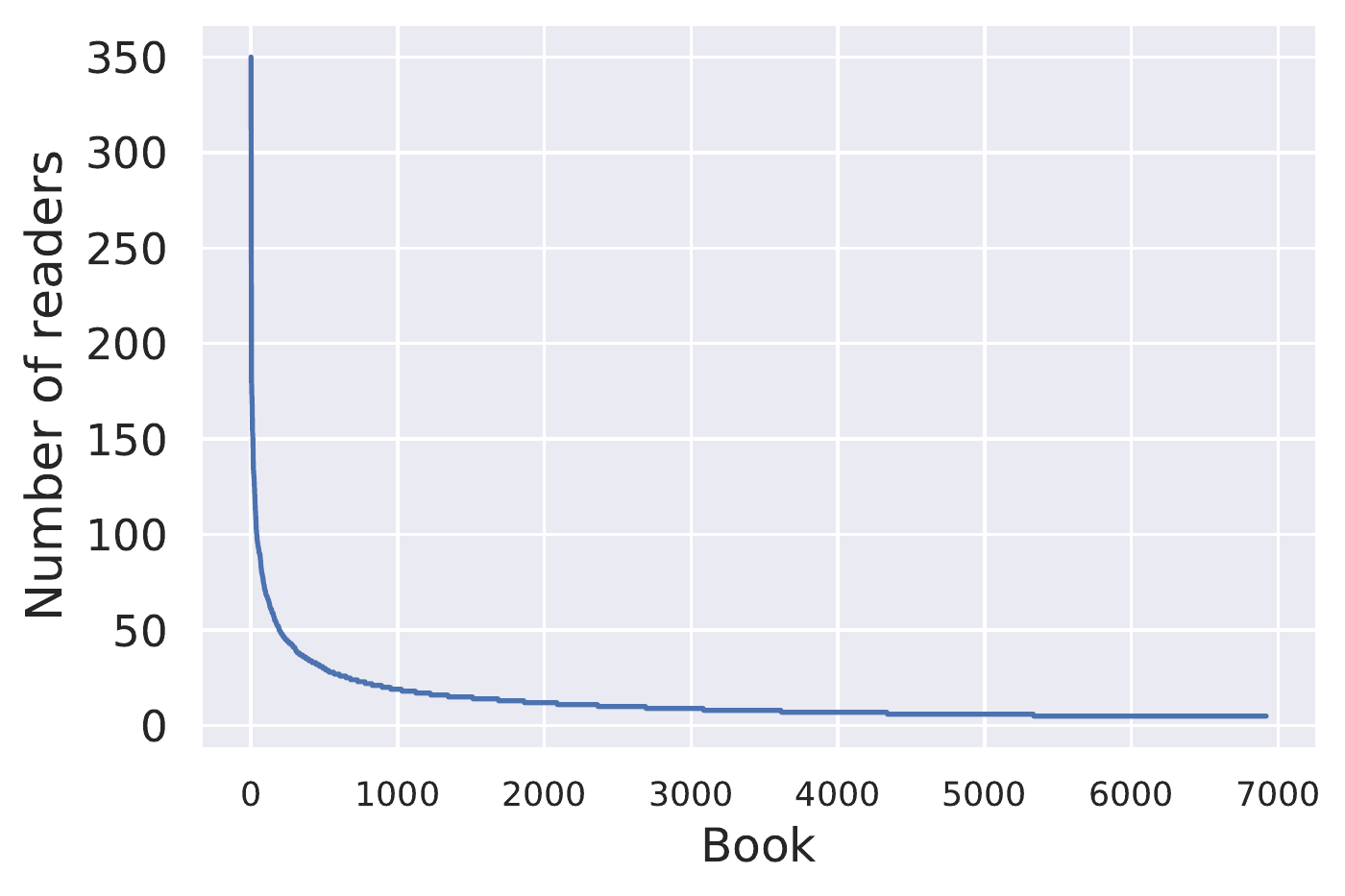}
            }
            \label{fig:long_tail}
        }
    \subfloat[The ratio of popular items in users' profiles]
        {
            {
                \includegraphics[width=5cm]{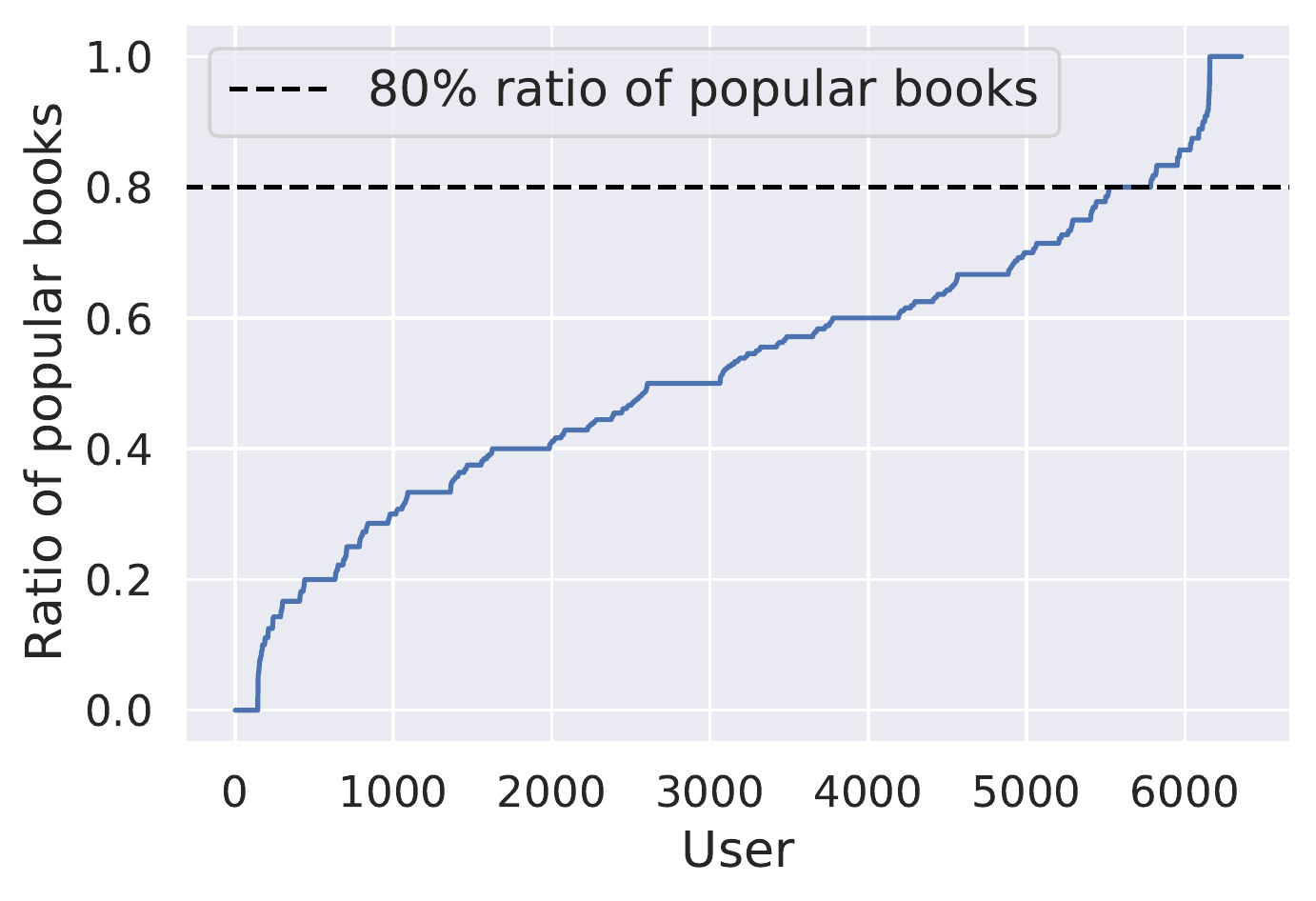} 
            }
            \label{fig:book_ratio}
        }
    \caption{Reading distribution of books.}
    \label{fig:distribution_of_books}
\end{figure}

\begin{itemize}
    \item \textbf{RQ1}: How much are different individuals or groups of users interested in popular books?
    \item \textbf{RQ2}: How does the popularity bias in recommendation algorithms impact users with different tendencies toward popular books?
\end{itemize}

In the following, we explore the \dataset dataset characteristics and investigate \textbf{RQ1} in Section \ref{sec:data_bias}. Then, we examine \textbf{RQ2} in Section \ref{sec:bookrec}. We discuss the findings of our study and conclude the paper in Sections \ref{sec:discussion} and \ref{sec:conclusion}, respectively. Finally, to enable reproducibility of the experiment, we have made our codes open source.\footnote{\url{https://github.com/rahmanidashti/FairBook}}

\section{Popularity Bias in Data}
\label{sec:data_bias}
In this paper, we utilize the well-known \textit{\dataset}\footnote{\url{http://www2.informatik.uni-freiburg.de/~cziegler/BX/}} dataset that contains $1,149,780$ anonymous explicit and implicit ratings of approximately $340,556$ books made by $105,283$ users in a $4$-week crawl (August - September $2004$) \cite{ziegler2005improving}. From the dataset, we first removed all the implicit ratings, then we removed users who had fewer than $5$ ratings so that the retained users were those who were likely to have rated enough long-tail items.
The limit of 5 ratings was also used to remove distant long-tail items. Once short-profile users and distant long-tail items are removed, the \dataset dataset consists of $6,358$ users who rated $6,921$ books, totaling $88,552$ in ratings. In the following, we will address the first research question. Table \ref{tbl:dataset_char} shows the characteristics of the \dataset datasets.

\subsection{Reading Distribution of Books}
\label{sec:reading_distribution_of_books}
Fig.~\ref{fig:distribution_of_books} illustrates the distribution of readings in \dataset dataset. Figure~\ref{fig:long_tail} indicates that reading counts of books follow a long-tail distribution as expected. 
That means a small proportion of books are read by many users, whereas a significant proportion (\ie the long-tail) is read by only a small number of readers. Additionally, we illustrate in Fig~\ref{fig:book_ratio} the ratio of popular books to all books read by users. Same as \cite{abdollahpouri2019unfairness}, we sort books based on the number of readers and consider them as popular if they are within the top-$20$\% of the sorted list. By this definition, we observe that around $5,256$ out of $6,358$ users (\ie around $83\%$ of users) have read at least $20$\% of unpopular books in their profile.

\begin{table}[t]
  \caption{Statistics of the \dataset dataset.}
  \centering
  \label{tbl:dataset_char}
  \begin{tabular}{cccccc}
    \toprule
    \#Users & \#Books & \#Interactions & $\frac{\#Interactions}{\#Users}$ & $\frac{\#Interactions}{\#Books}$ & Sparsity \\
    \midrule
    6,358 & 6,921 & 88,552 & 13.92 & 12.79 & 99.80\% \\
    \bottomrule 
  \end{tabular}
\end{table}

\begin{figure}[t]
    \centering
    \subfloat[Correlation of profile size and the number of popular items in user profile]
        {
            {
                \includegraphics[scale=0.36]{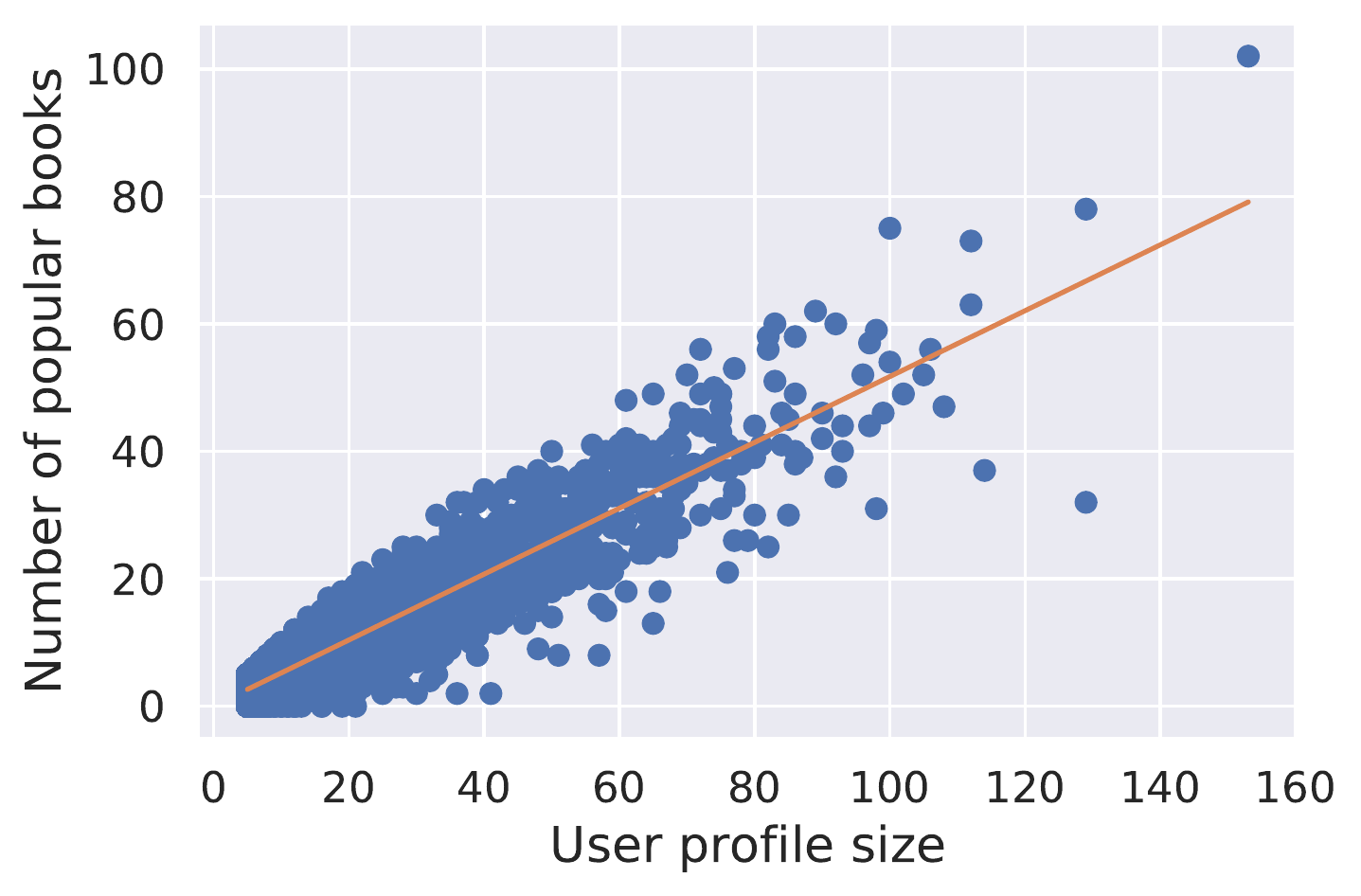}
            }
            \label{fig:corr_user_pop}
        }
    \qquad
    \subfloat[Correlation of profile size and the average popularity of items in user profile]
        {
            {
                \includegraphics[scale=0.36]{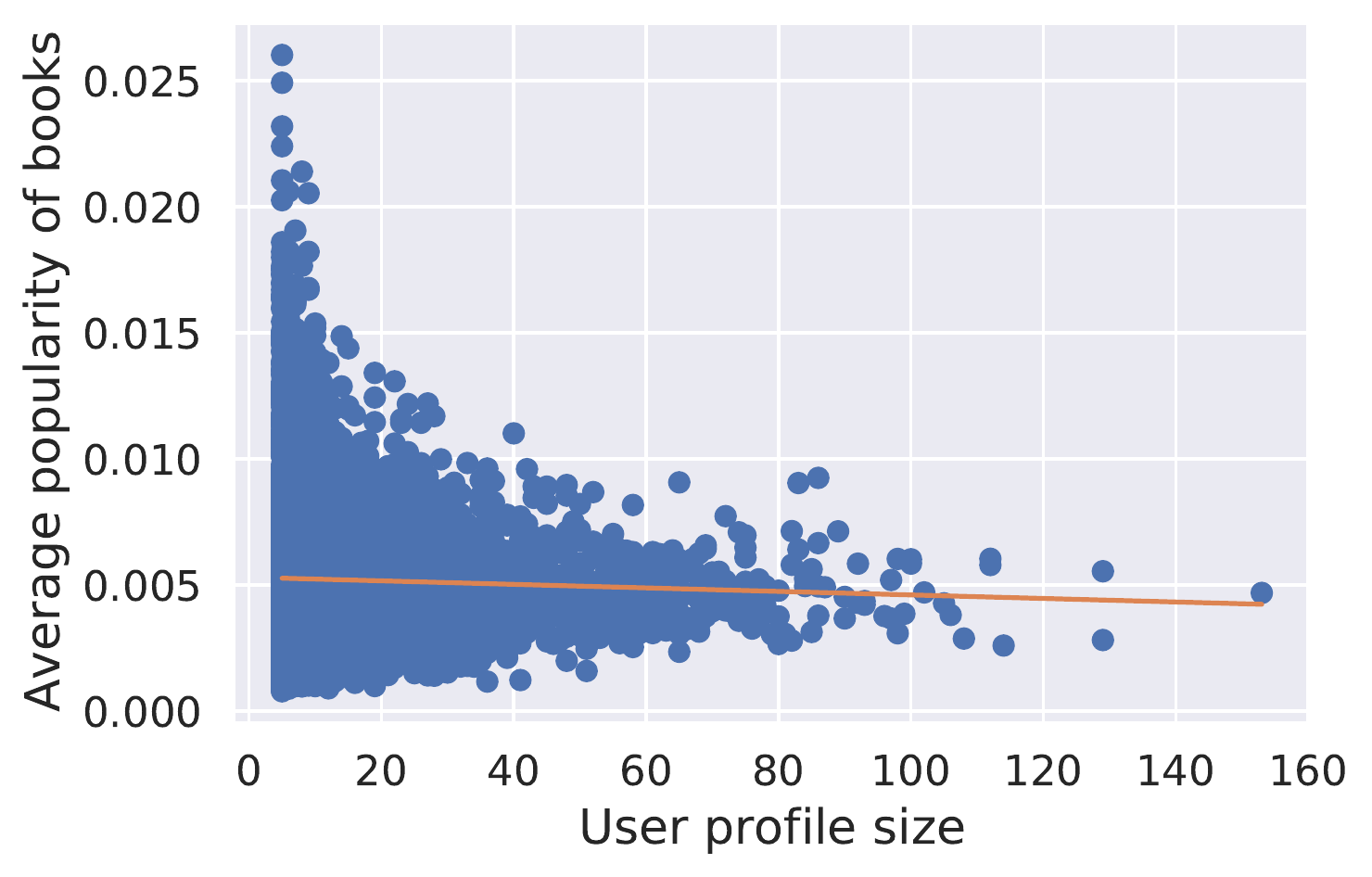} 
            }
            \label{fig:corr_user_avg}
        }
    \caption{Correlation of user profile size and the popularity of books in the user profile. While there is a positive correlation between profile size and number of popular books, there is a negative correlation between profile size and the average book popularity.}
    \label{fig:correlation}
\end{figure}

\subsection{User Profile Size and Popularity Bias in Book Data} \label{user_profile_size_and_popularity_bias_in_book_data}
In Fig.~\ref{fig:correlation} we investigate whether a correlation exists between the size of the user profile and the presence of popular books in the profile. Specifically, in Fig.~\ref{fig:corr_user_pop}, we depict the number of popular books in a user's profile over the size of the profile. As could be expected, there is a positive correlation since the more items in a user profile, the greater probability there are popular items in the profile. On the other hand, when plotting the average popularity of books in a user profile over the profile size in Fig.~\ref{fig:corr_user_avg}, we observe a negative correlation, which indicates that users having a smaller profile size tend to read books with higher average popularity. Same as \cite{abdollahpouri2019unfairness}, we define the popularity of a book as the ratio of users who have read that book.

As stated before, in this work, we investigate the popularity bias from the users' perspective, same as \cite{abdollahpouri2019unfairness}. To this end, we categorized all users into three groups according to their profile's ratio of popular items (\ie book). This ratio indicates how interested they are in popular items.

\begin{itemize}
     \item \textbf{Niche users}: After sorting users based on the ratio of popular items in their profiles, we refer to the bottom $20\%$ of the sorted list as Niche users.
     \item \textbf{Bestseller-focused users}: We consider the top $20\%$ users of the sorted list as Bestseller-focused users interested in popular books.
     \item \textbf{Diverse users}: The rest of the users fall within this category. Users in this category have varied interests in popular and unpopular books.
\end{itemize}

\begin{figure}[t]
    \centering
    \includegraphics[scale=0.45]{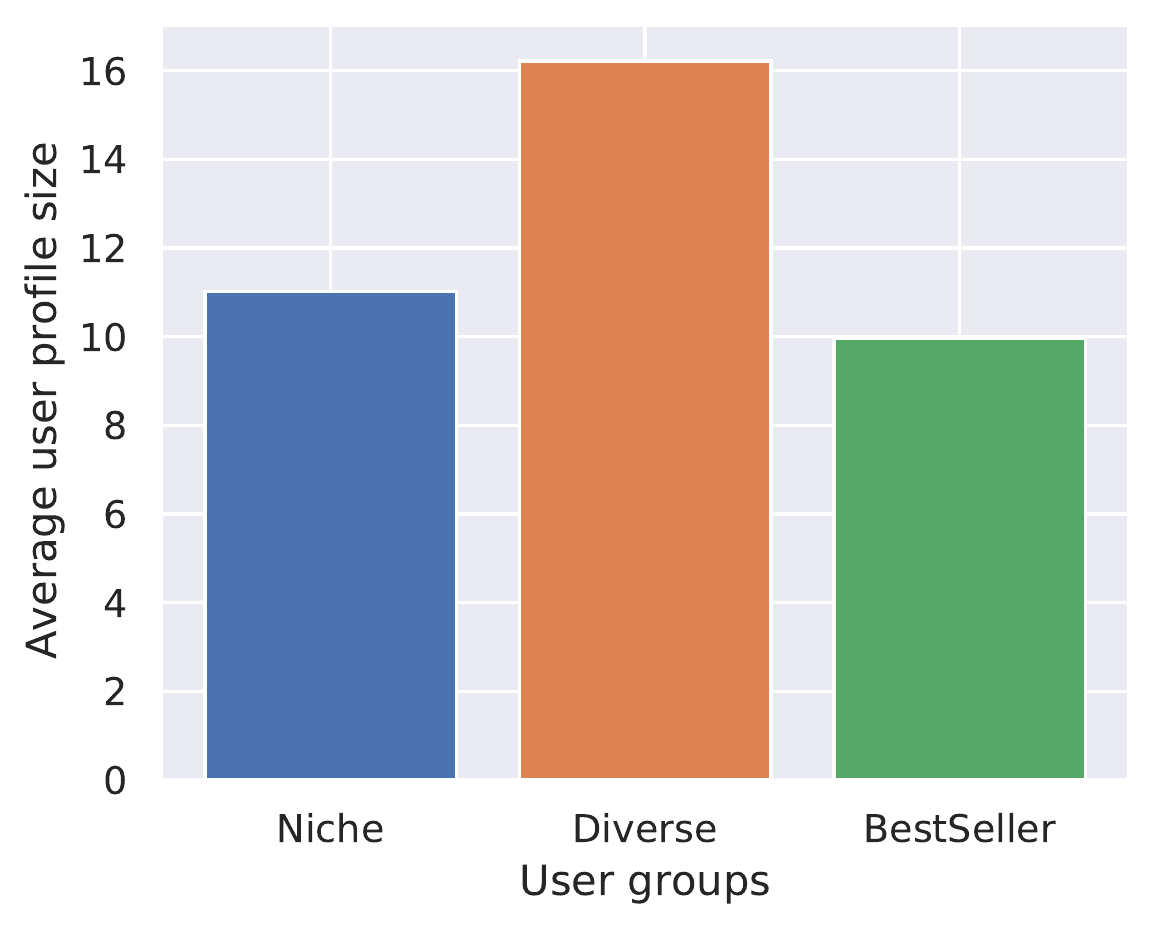}
    \caption{The average profile size in different user groups}
    \label{fig:group_size}
\end{figure}

Fig.~\ref{fig:group_size} indicates the average profile size of different user groups. As expected, Diverse users have the largest profile size, followed by Niche users. The Bestseller-focused group has the smallest average profile size. Based on our analysis in section \ref{user_profile_size_and_popularity_bias_in_book_data}, diverse users have larger average profile size; therefore, we can expect them to read more popular books than niche users. Furthermore, the small profile size of Bestseller-focused users implies that they are focused on reading solely popular books.
Finally, we expect that recommendation algorithms influenced by popularity bias will provide Bestseller-focused users with the best recommendation quality (\ie accuracy), followed by Diverse users. Furthermore, Niche users are likely to receive the lowest recommendation quality, as they have the lowest ratio of popular items in their profile. We will explore these expectations in the next section.

Hence, in this section, we find that majority of users (\ie around five-seventh) have read at least $20$\% of unpopular books. Furthermore, we find that users with a small profile size tend to read more popular books than users having a larger profile size. To sum up, we investigated \textbf{RQ1} in this section and our findings are in agreement with what Abdollahpouri et al.~have reported in \cite{abdollahpouri2019unfairness}.

\begin{figure}[htp]
    \centering
    \subfloat[Random]
        {
            {
                \includegraphics[width=3.7cm]{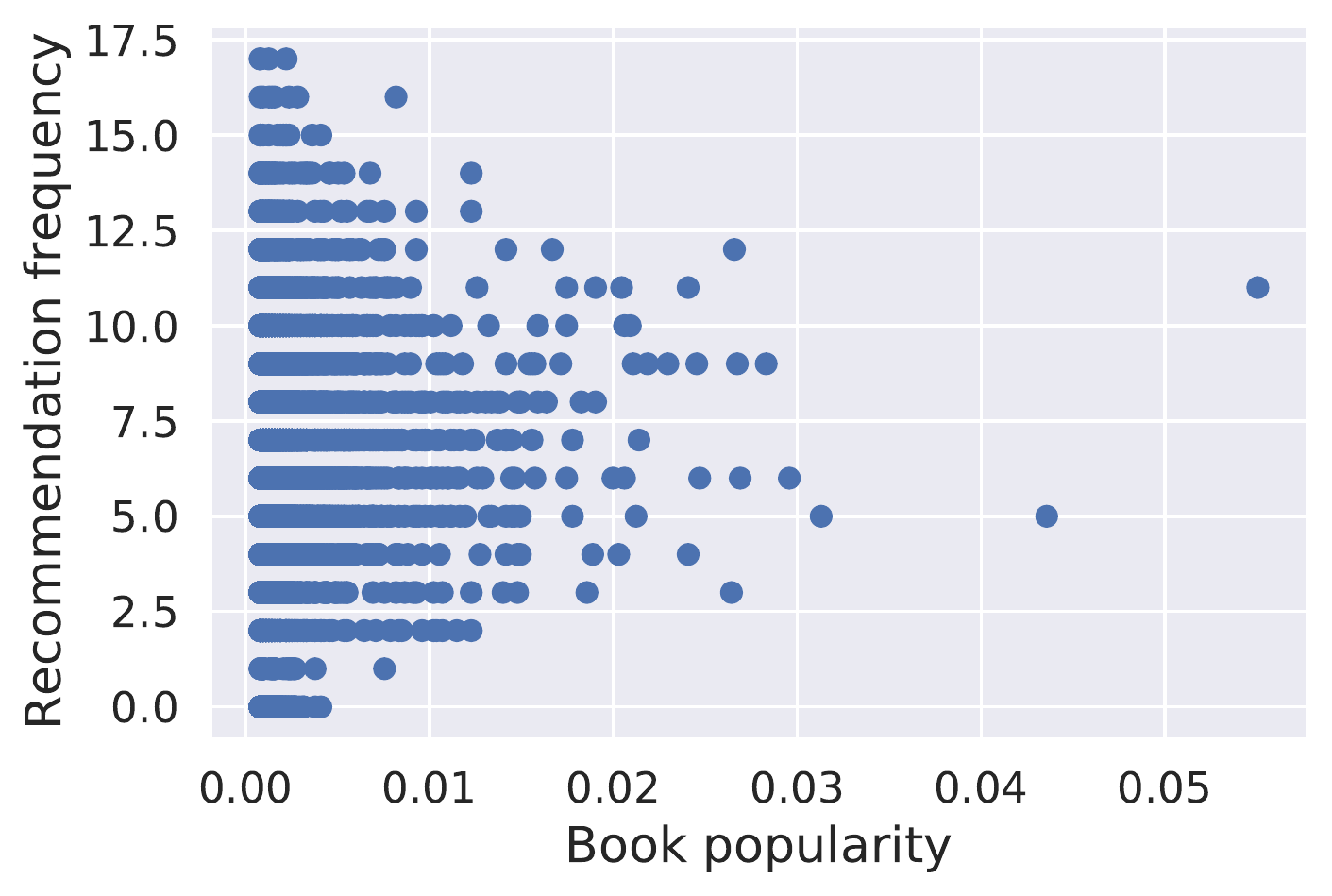}
            }
        }
    \subfloat[MostPop]
        {
            {
                \includegraphics[width=3.7cm]{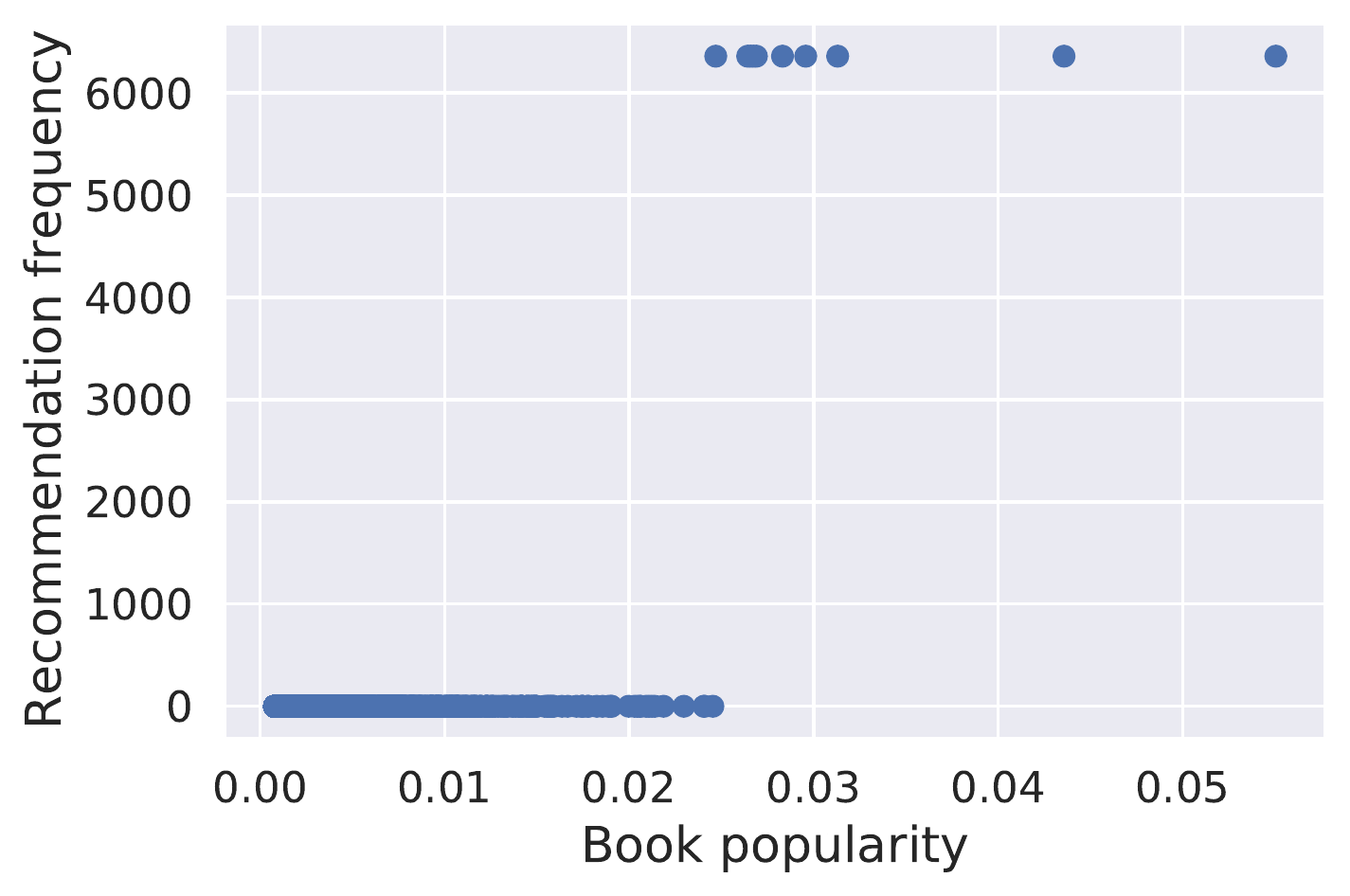} 
            }
        }
    \subfloat[UserKNN]
        {
            {
                \includegraphics[width=3.7cm]{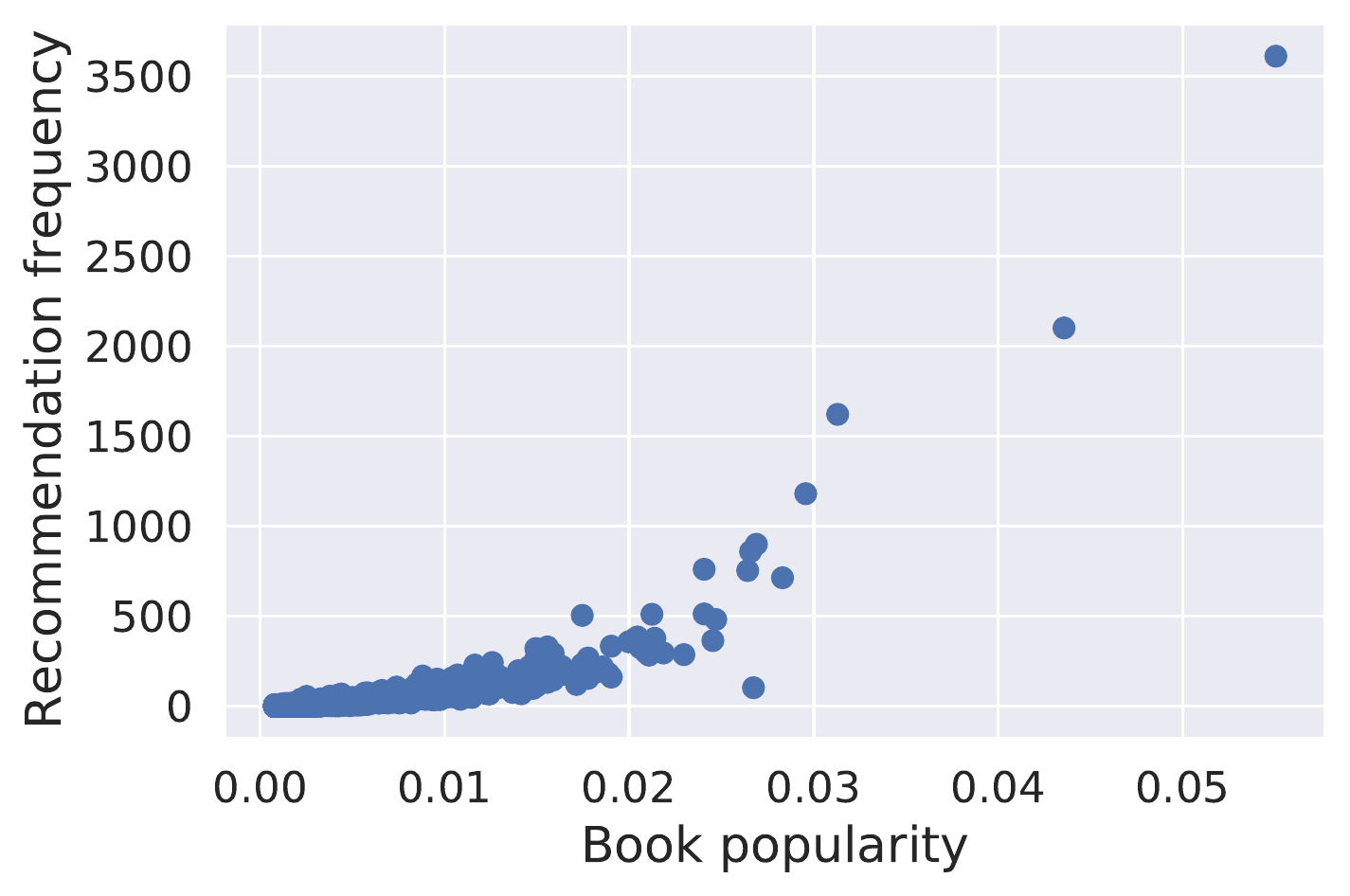} 
            }
        }
    \qquad
    \subfloat[MF]
        {
            {
                \includegraphics[width=3.7cm]{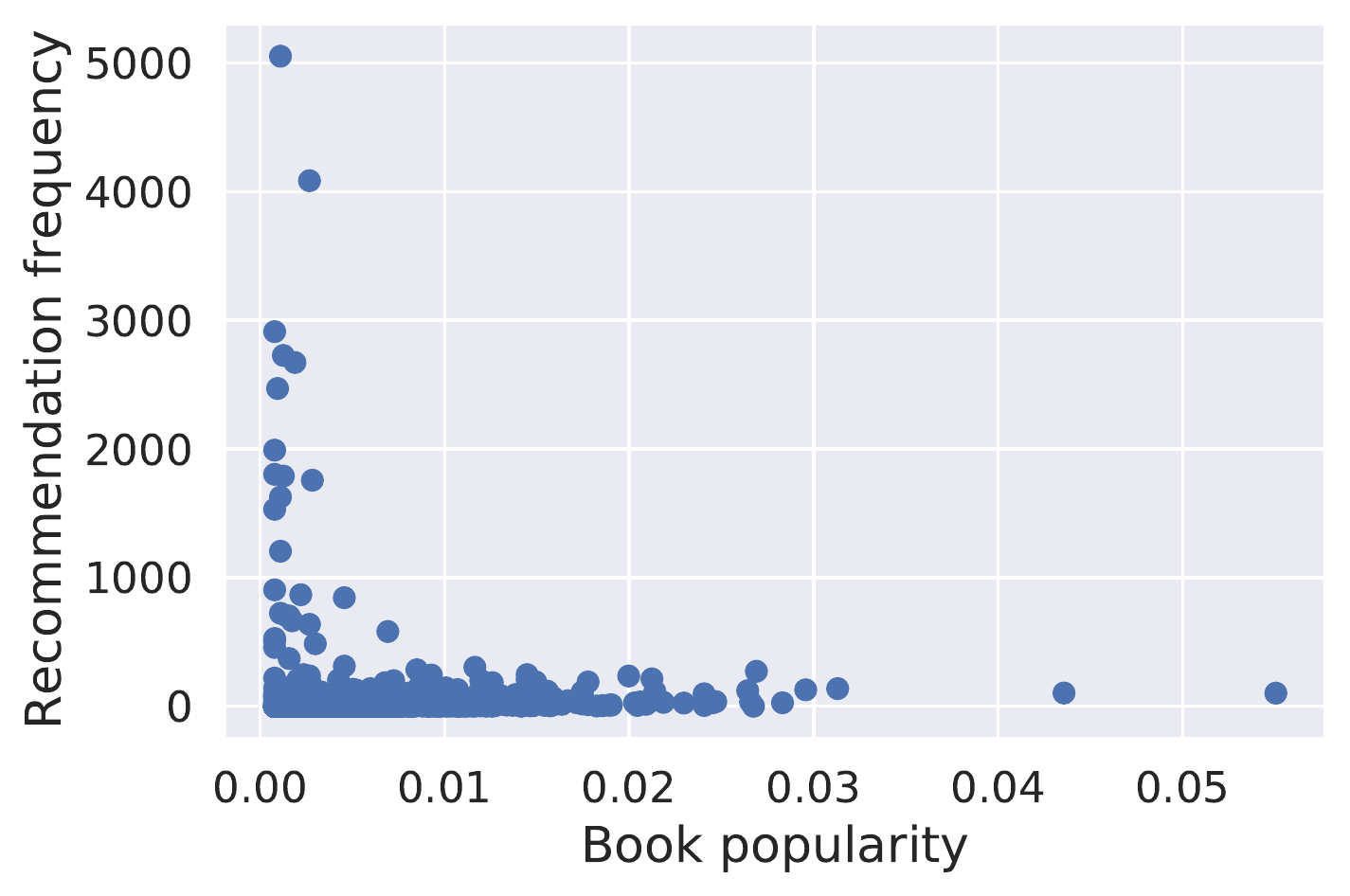} 
            }
        }
    \subfloat[PMF]
        {
            {
                \includegraphics[width=3.7cm]{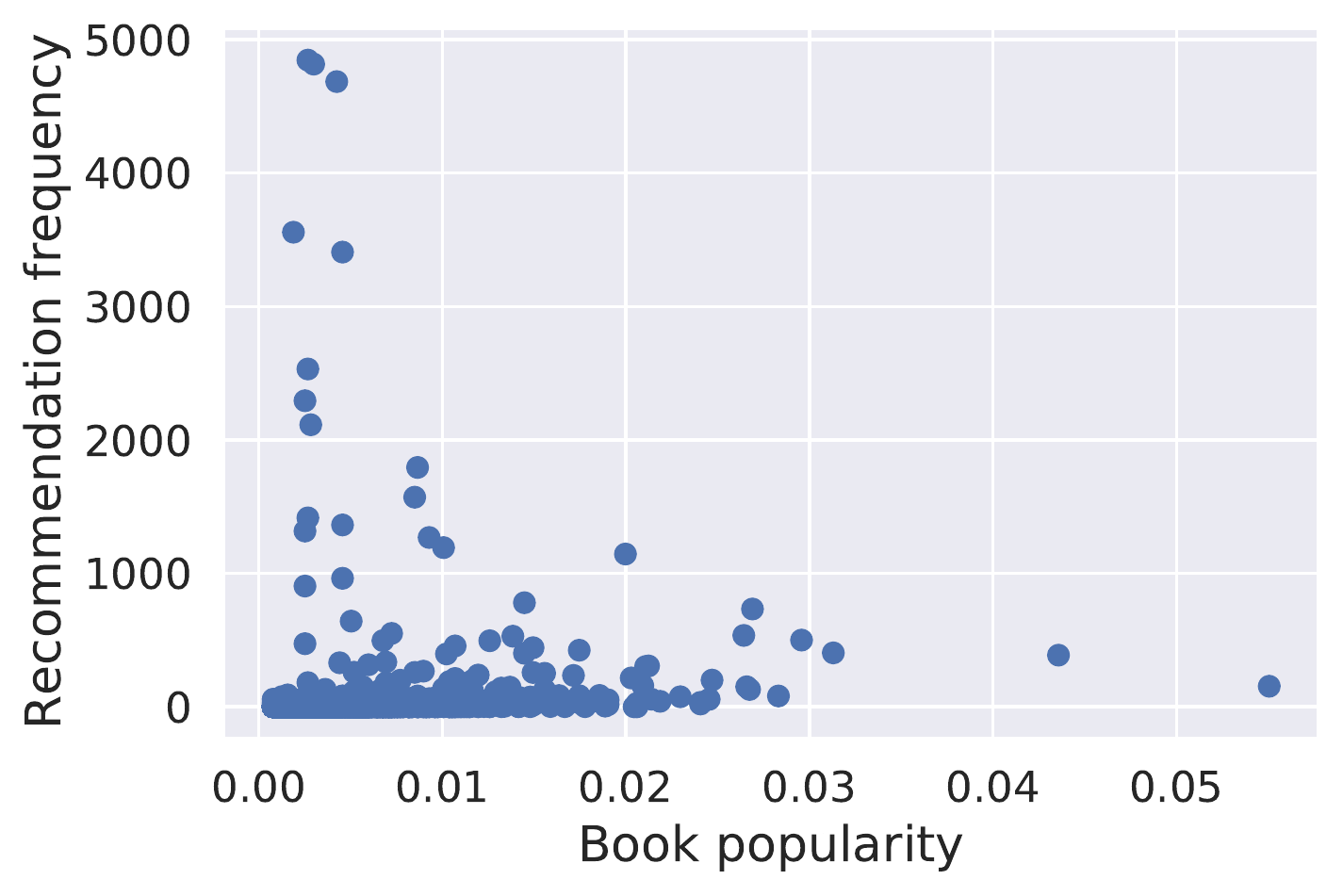} 
            }
        }
    \subfloat[NMF]
        {
            {
                \includegraphics[width=3.7cm]{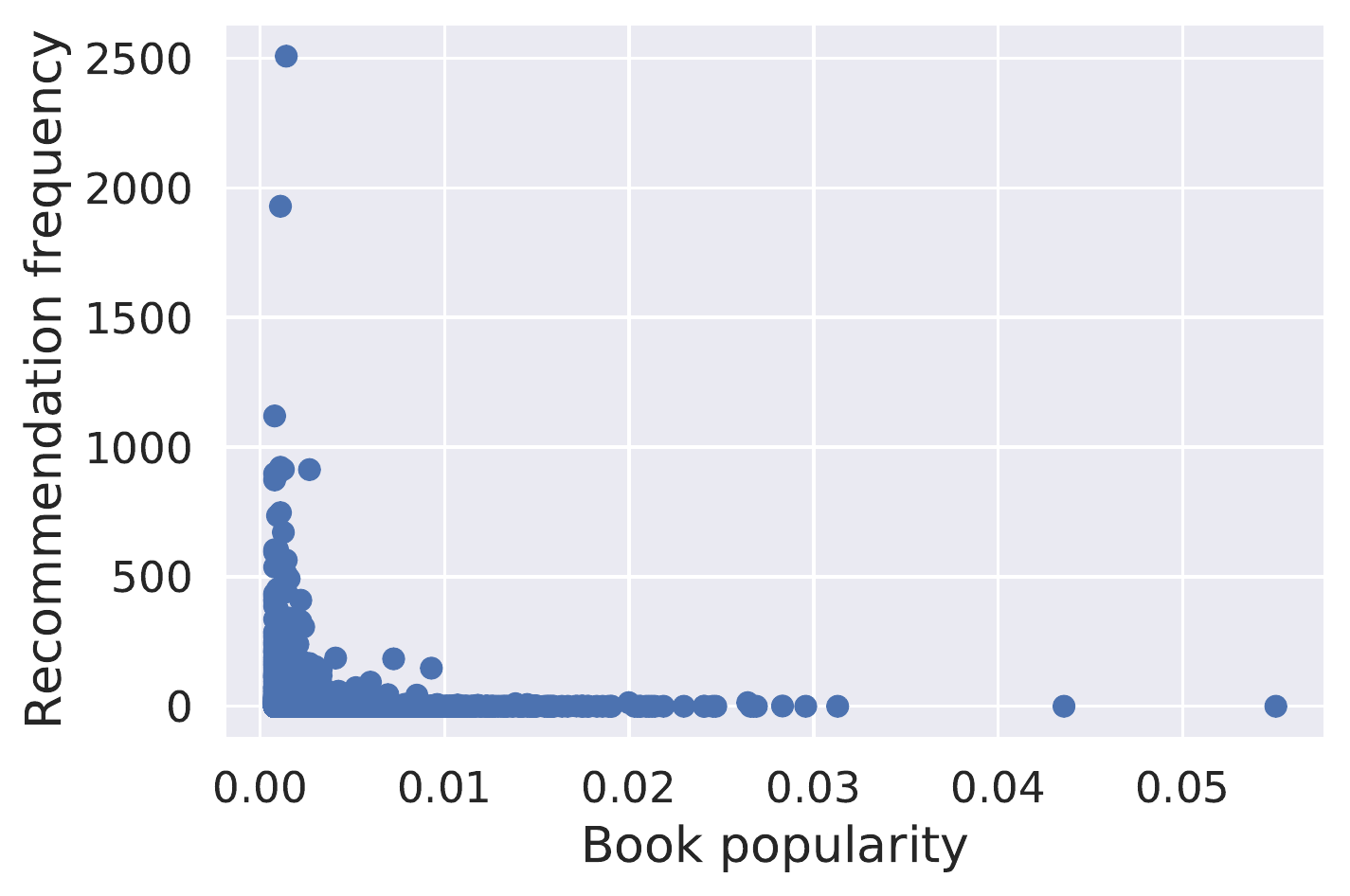} 
            }
        }
    \qquad
    \subfloat[WMF]
        {
            {
                \includegraphics[width=3.7cm]{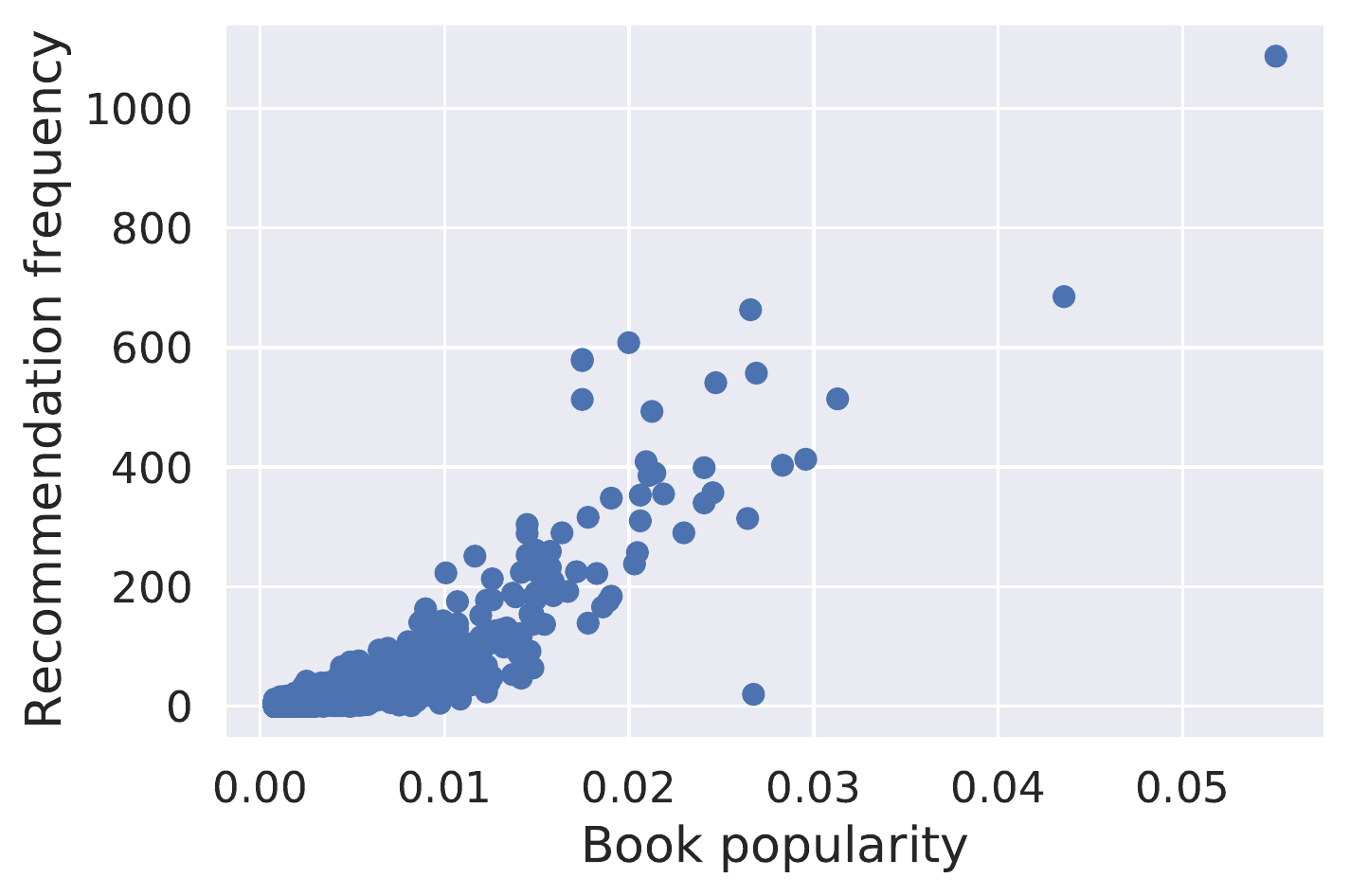} 
            }
        }
    \subfloat[BPR]
        {
            {
                \includegraphics[width=3.7cm]{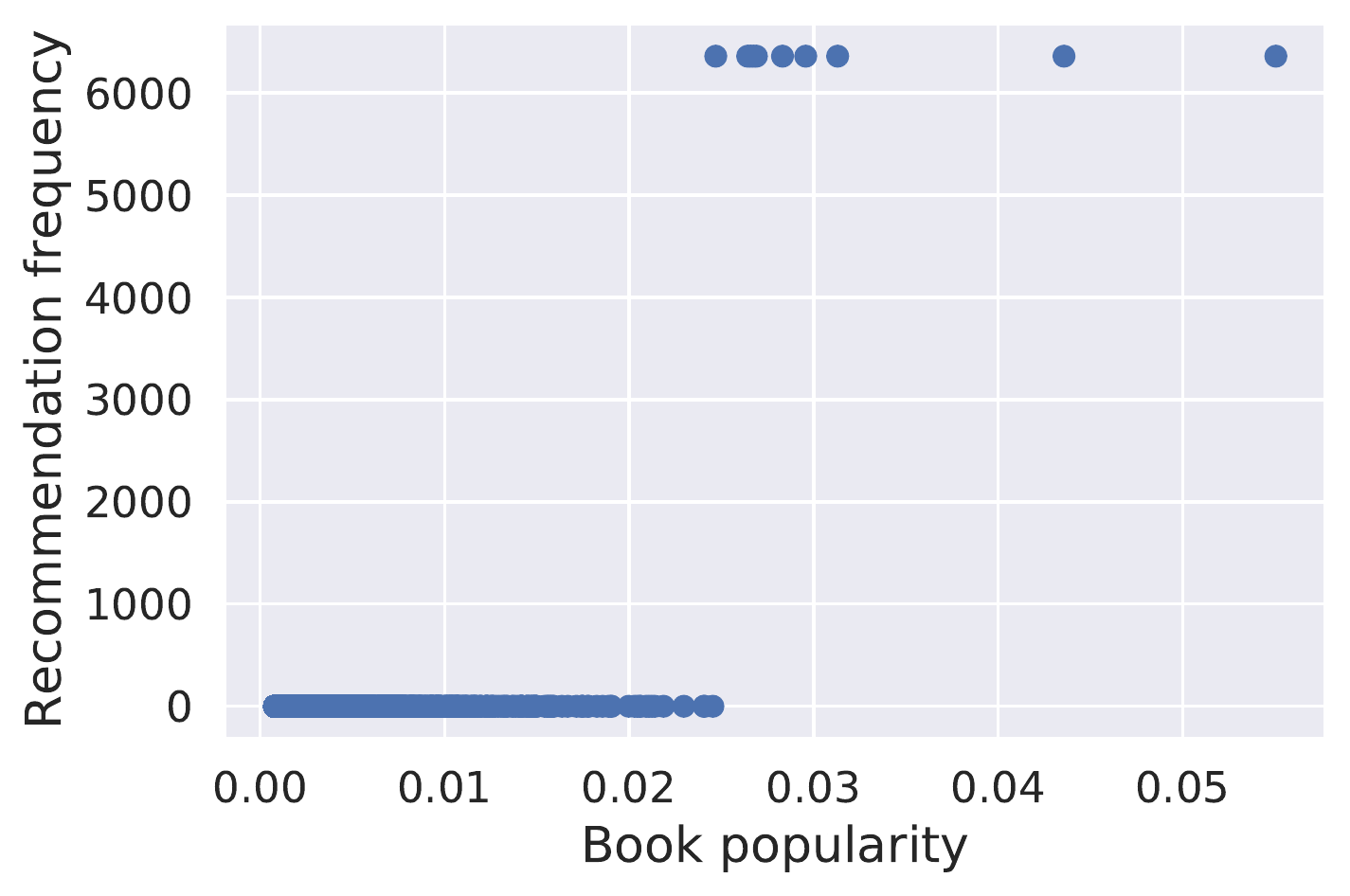} 
            }
        }
    \subfloat[PF]
        {
            {
                \includegraphics[width=3.7cm]{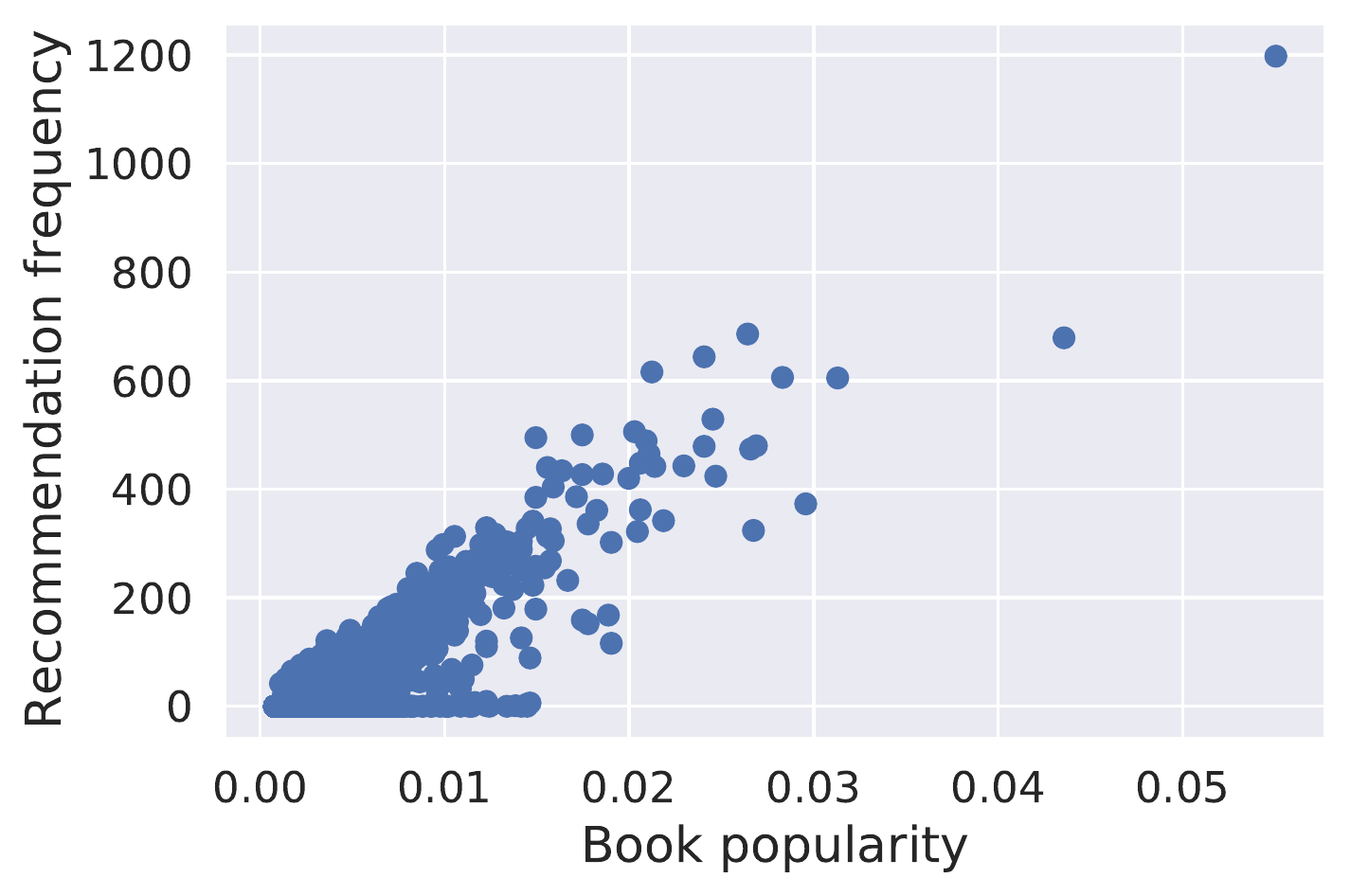} 
            }
        }
    \qquad
    \subfloat[NeuMF]
        {
            {
                \includegraphics[width=3.7cm]{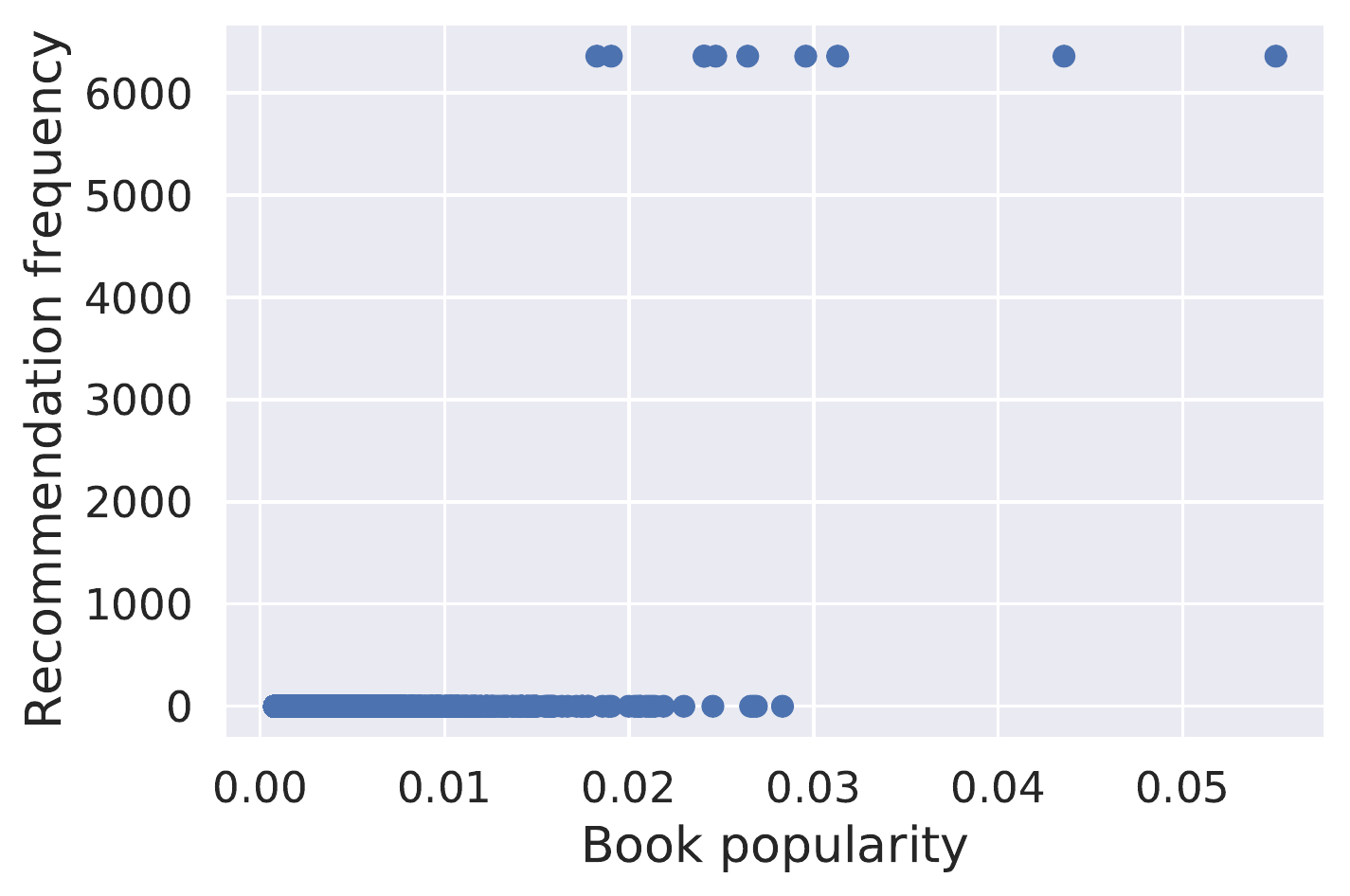} 
            }
        }
    \qquad
    \subfloat[VAECF]
        {
            {
                \includegraphics[width=3.7cm]{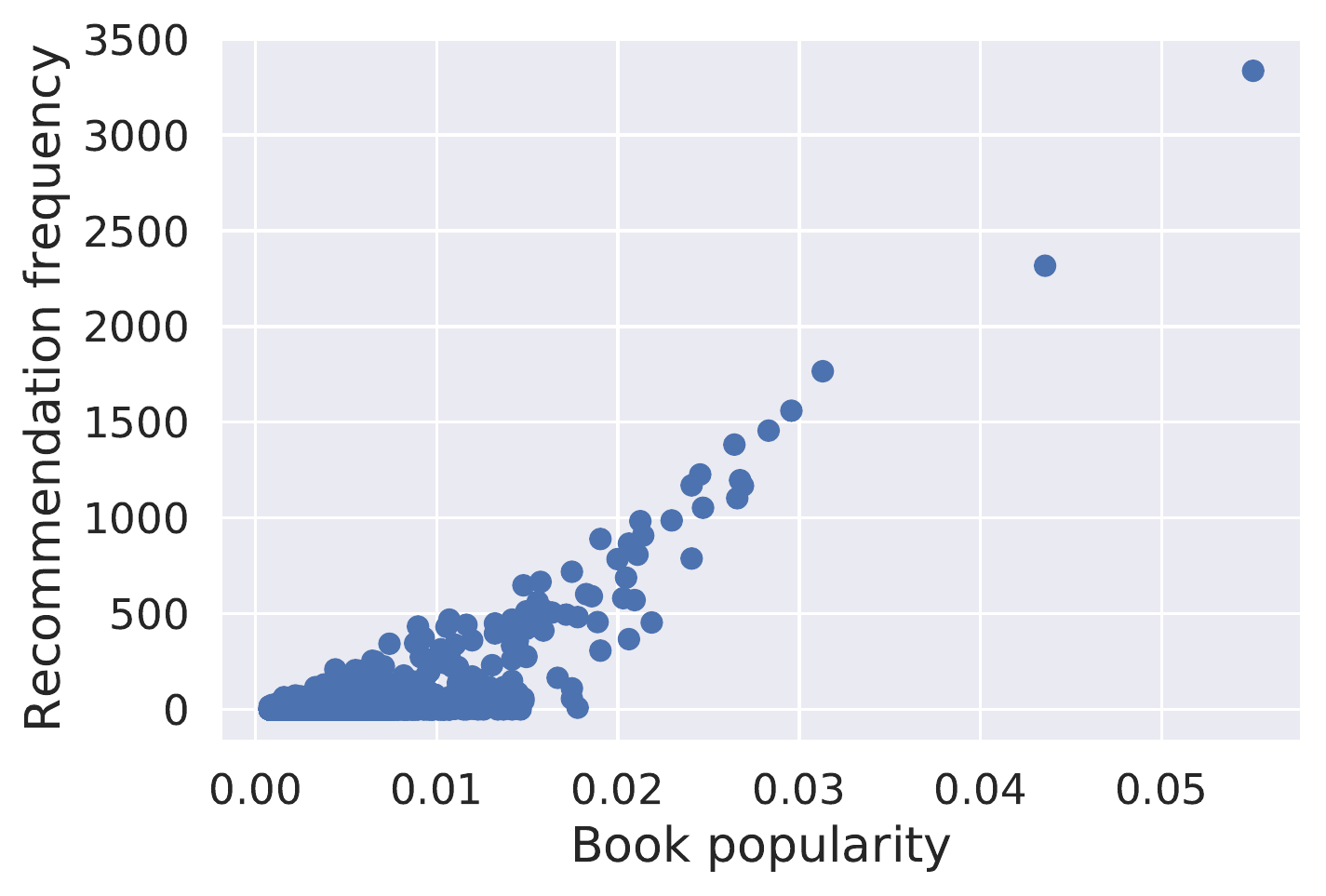} 
            }
        }
    \caption{The correlation between the popularity score of items and the number of times they are being recommended using different algorithms}
    \label{fig:alorithm_correlation}
\end{figure}

\section{Popularity Bias in Book Recommendation}
\label{sec:bookrec}
In this section, we study popularity bias in state-of-the-art book recommendation algorithms. In order to foster the reproducibility of this study, we implement and evaluate all recommendation algorithms using Cornac\footnote{\url{https://cornac.preferred.ai/}}, which is a Python-based open source recommendation framework \cite{salah2020cornac,truong2021exploring}. Therefore, we formulate the book recommendation using Cornac as a rating prediction problem, where we predict the preference of a target user $u$ for a target book $b$. Then, we recommend the top-$10$ books with the highest predicted preferences.

For ease of comparison, we use the same evaluation setup used in \cite{abdollahpouri2019unfairness} to evaluate the performance of the recommendation algorithms. To this end, we set aside $80\%$ of the \dataset dataset as training set and the remaining 20\% as the test set. We further extended the prior study of \cite{abdollahpouri2019unfairness} by incorporating a more comprehensive range of state-of-the-art algorithms, including
(i) baseline approaches, (ii) K-Nearest Neighbour (KNN) approaches, (iii) Matrix Factorization (MF) approaches, and (iv) Neural Network (NN) approaches. Specifically, we evaluate two baselines approaches, \ie \ran and \topp. We include the \uknn \cite{breese2013empirical}, which is a KNN-based method. We also evaluate five MF-based approaches, including \mf \cite{koren2009matrix}, \pmf \cite{mnih2008probabilistic}, \nmf \cite{lee2000nmf}, \wmf \cite{hu2008collaborative}, and \pf \cite{gopalan2015scalable}. We also consider \bpr \cite{rendle2009bpr} as one of the well-know ranking-based algorithms. Eventually, we include two state-of-the-art NN-based approaches, \ie \neumf \cite{he2017neural} and \vaecf \cite{liang2018variational}. We adopt the recommendation algorithm with the default hyperparameter settings suggested in their original paper.

\subsection{Recommendation of Popular Books}
\label{sec:recommendation}
As shown in our analysis in Section \ref{sec:data_bias}, there is an imbalanced distribution in the book rating data, \ie certain books are rated very frequently while the majority of items are rated by only a few users. In this section, we investigate to what extent different recommendation algorithms propagate this bias into their recommendations. First, we examine algorithms' overall performance without taking into account how they perform for different users or groups of users based on their tendency towards popular items. In Fig.~\ref{fig:alorithm_correlation}, we illustrate the correlation of book popularity and how often the eight algorithms recommend these books. Among baseline algorithms, we are seeing a strong positive correlation on \topp showing algorithm tendency to recommend popular items frequently not giving chance to long-tail items and no meaningful correlation on \ran as expected. Interestingly, we observe a strong correlation between the popularity of items and their recommendation frequency on \neumf and \bpr with very similar behavior to \topp. A majority of books were not exposed to users by these algorithms, while popular ones are more frequently highlighted.
Furthermore, our experiment shows a moderate positive correlation in \wmf and \pf among the Matrix Factorization-based approaches. In contrast to Abdollahpouri et al.~\cite{abdollahpouri2019unfairness} and Kowald et al.~\cite{kowald2020unfairness} study in the Movie and Music domain, no positive correlation exists in \pmf, \mf, and \nmf, indicating that the latter algorithms in Matrix Factorization-based approaches are not prone to popularity bias in \dataset dataset. Additionally, this suggests that the characteristics of underlying data and the domain could play a key role in determining how recommendation algorithms behave in propagating popularity bias in various domains. Finally, among NN-based state-of-the-art approaches investigated in this study, the \vaecf model has a moderately positive correlation and a better performance than the \neumf model.

\begin{figure}[t]
    \centering
    \includegraphics[scale=0.5]{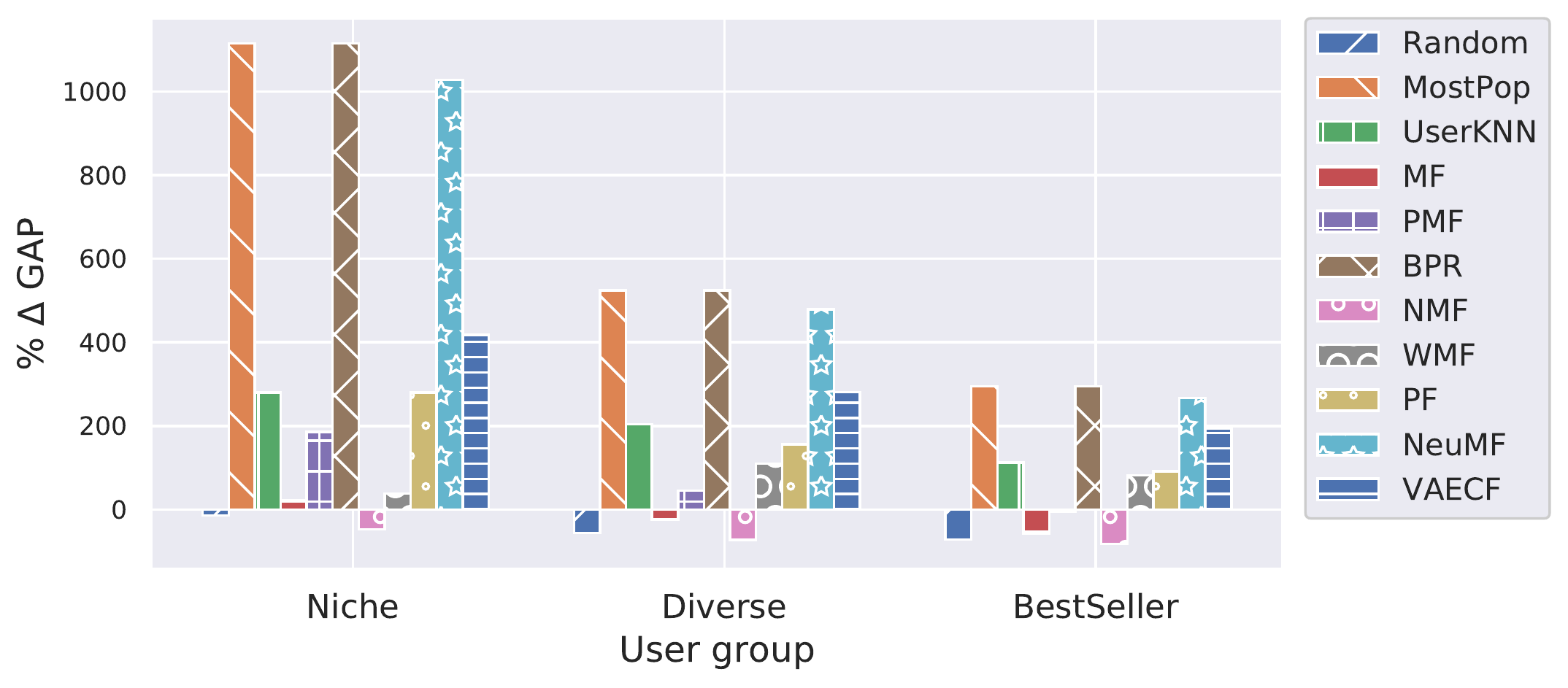}
    \caption{The Group Average Popularity ($\Delta$GAP) of different algorithms for Niche, Diverse, and Bestseller-focused user groups. Except for the \ran, \mf, \nmf, \pmf, and \wmf algorithms, all approaches provide too popular book recommendations for all three user groups.}
    \label{fig:gap_analysis}
\end{figure}

\begin{figure}
    \centering
    \subfloat[MAE]
        {
            {
                \includegraphics[scale=0.38]{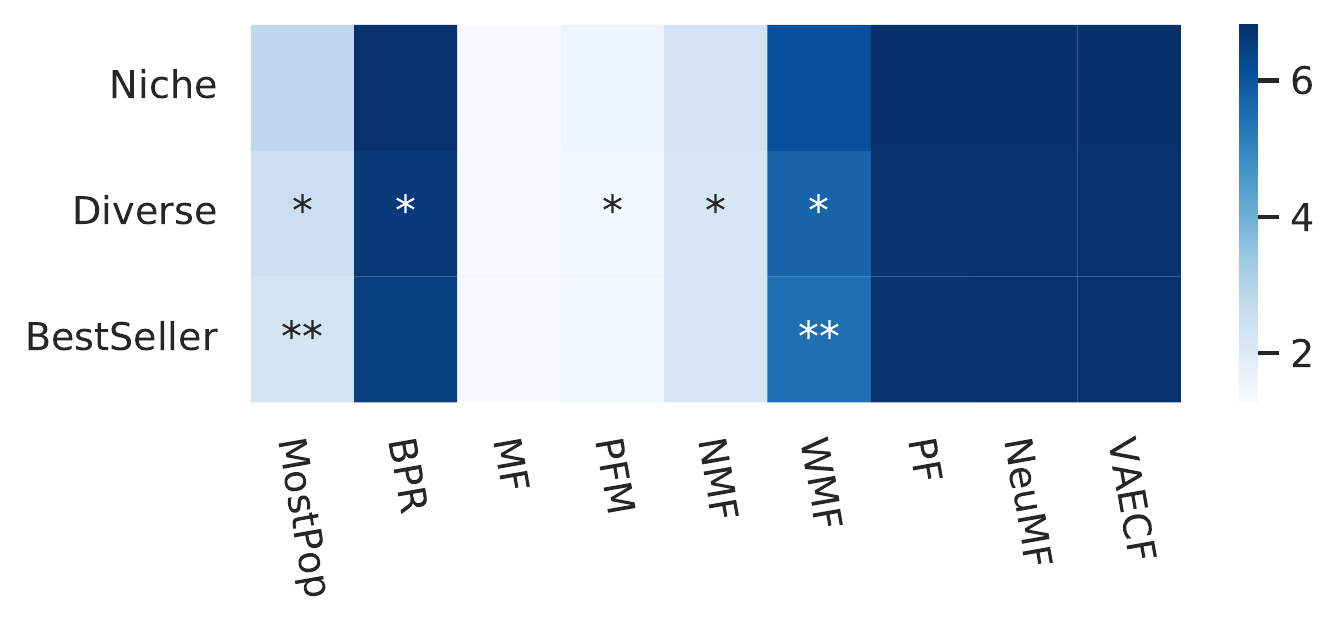}
            }
        }
    \subfloat[Precision]
        {
            {
                \includegraphics[scale=0.38]{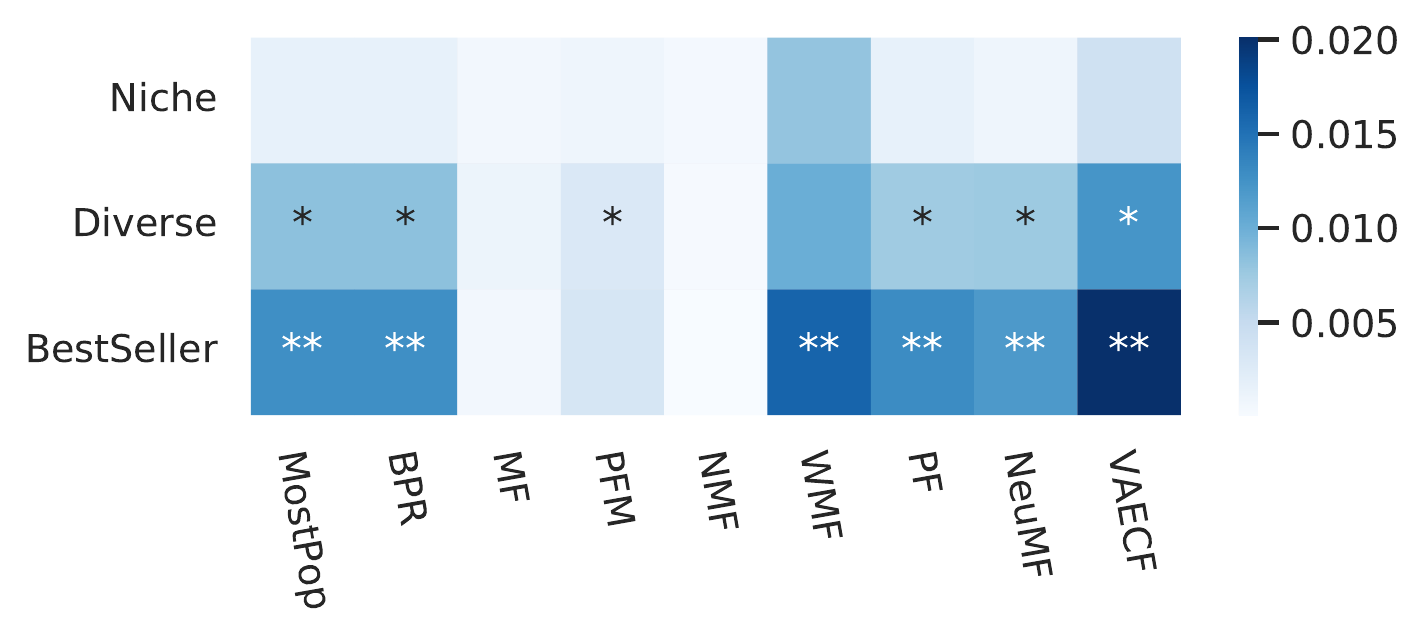} 
            }
        }
    \qquad
    \subfloat[Recall]
        {
            {
                \includegraphics[scale=0.38]{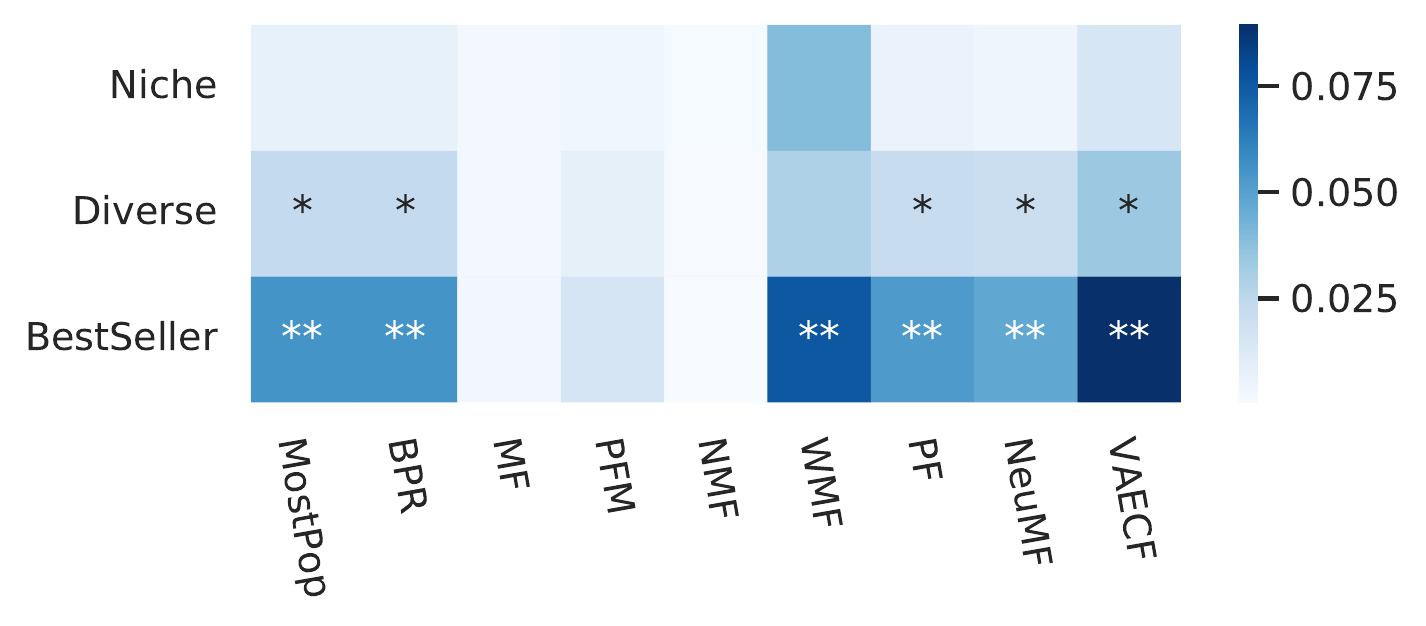} 
            }
        }
    \subfloat[NDCG]
        {
            {
                \includegraphics[scale=0.38]{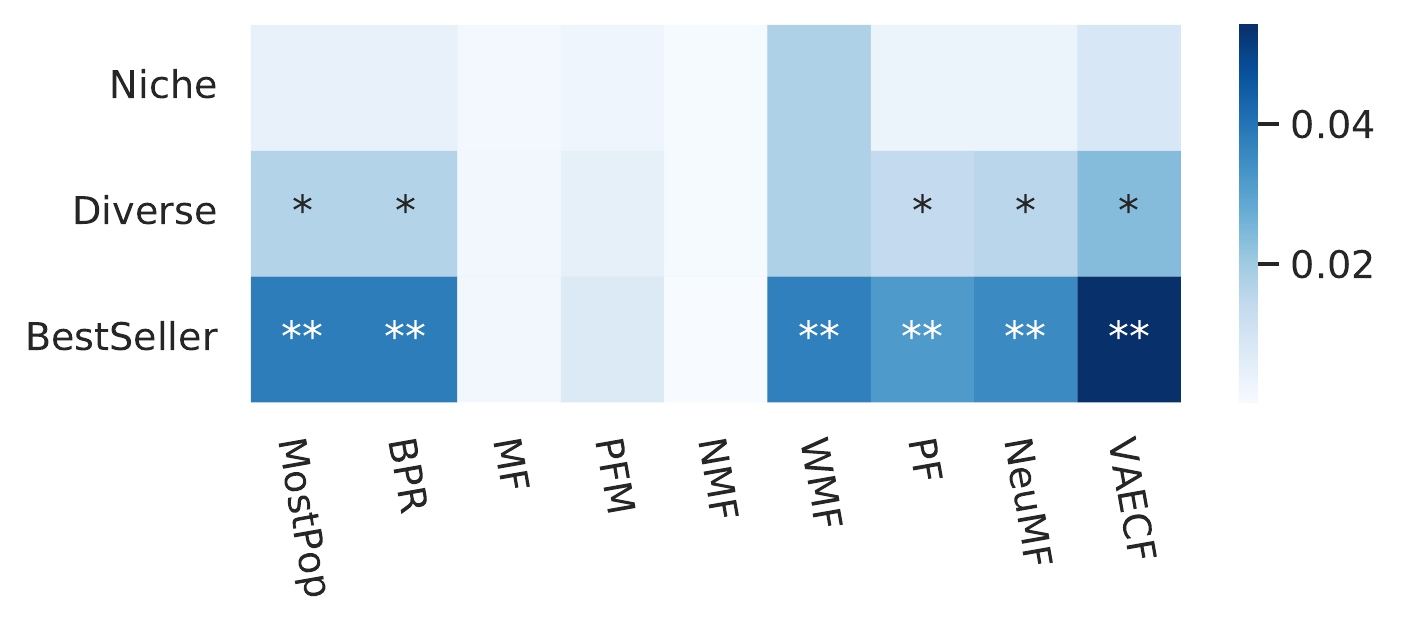} 
            }
        }
    \caption{The performance of models for the three user groups in terms of MAE (the lower, the better) and Precision, Recall, and NDCG (the higher, the better). The best results are always given for the \textbf{Bestseller-focused} user group (statistically significant according to a t-test with $p < 0.005$ as indicated by **). Across the algorithms, the best results are provided by \textbf{VAECF} (indicated by dark blue colour).}
    \label{fig:groups_results}
\end{figure}

\subsection{Popularity Bias for Different User Groups}
\label{sec:group_bias}
In this section, we investigate the effect of popularity bias on different user groups (\ie Niche, Diverse, and Bestseller-focused) in book recommendations. To this end, we use \textit{delta Group Average Popularity} ($\Delta$GAP) metric proposed by Abdollahpouri et al.~\cite{abdollahpouri2019unfairness} to evaluate the unfairness of popularity bias. We further use \textit{MAE}, \textit{Precision}, \textit{Recall}, and \textit{NDCG} to evaluate the performance of recommender algorithms. For each recommendation algorithm and user group, $\Delta$GAP measures the difference between the average popularity of the recommended books and the average expected popularity shown in the user's profiles history as follows: 

\begin{equation}
    \Delta GAP=\frac{GAP(g)_r-GAP(g)_p}{GAP(g)_p}
\end{equation}

\noindent
where $g$ is a certain user group (\ie Niche, Diverse, or Bestseller-focused), $GAP(g)_r$ and $GAP(g)_p$ represent the GAP value for recommendation lists and user profiles, respectively, and it is defined as follows: 

\begin{equation}
    GAP(g)=\frac{\sum_{u\in{g}}\frac{\sum_{i\in{p_u}}{\phi{(i)}}}{|p_u|}^{}}{|g|}
\end{equation}

\noindent
where $\phi{(i)}$ is the popularity of item $i$, which is the number of times it is rated divided by the total number of users, and $p_u$ is the list of items in the profile of user $u$. The value of $\Delta$GAP = 0 indicates a fair recommendation meaning that the average popularity of recommended books matches the average popularity of the user profile.

Fig.~\ref{fig:gap_analysis} shows $\Delta$GAP values for recommendation algorithms among the three user groups. As can be seen, Niches users receive significantly higher average $\Delta$GAP values followed by Diverse and Bestseller-focused users, respectively. This finding confirms the results of the Abdollahpouri et al.~\cite{abdollahpouri2019unfairness} study and suggests that the popularity bias affects Niche users the most, that is, despite being interested in unpopular items, they receive recommendations of popular items. Interestingly, although the Bestseller-focused group receives the most favorable recommendations, the average $\Delta$GAP is $126.55$, revealing how algorithms can propagate the popularity bias even further than the Bestseller-focused user groups' interest in popular items. Fig. ~\ref{fig:gap_analysis} further illustrates that algorithms investigated in this study show similar behavior in terms of popularity bias among different user groups. Moreover, in line with our analysis in section~\ref{sec:recommendation}, \ran, \mf, and \nmf models provide the fairest recommendations, while \topp, \bpr, and \neumf suffer from the propagation of popularity bias across all user groups.

\begin{figure}[t]
    \centering
    \subfloat[Niche]
        {
            {
                \includegraphics[scale=0.37]{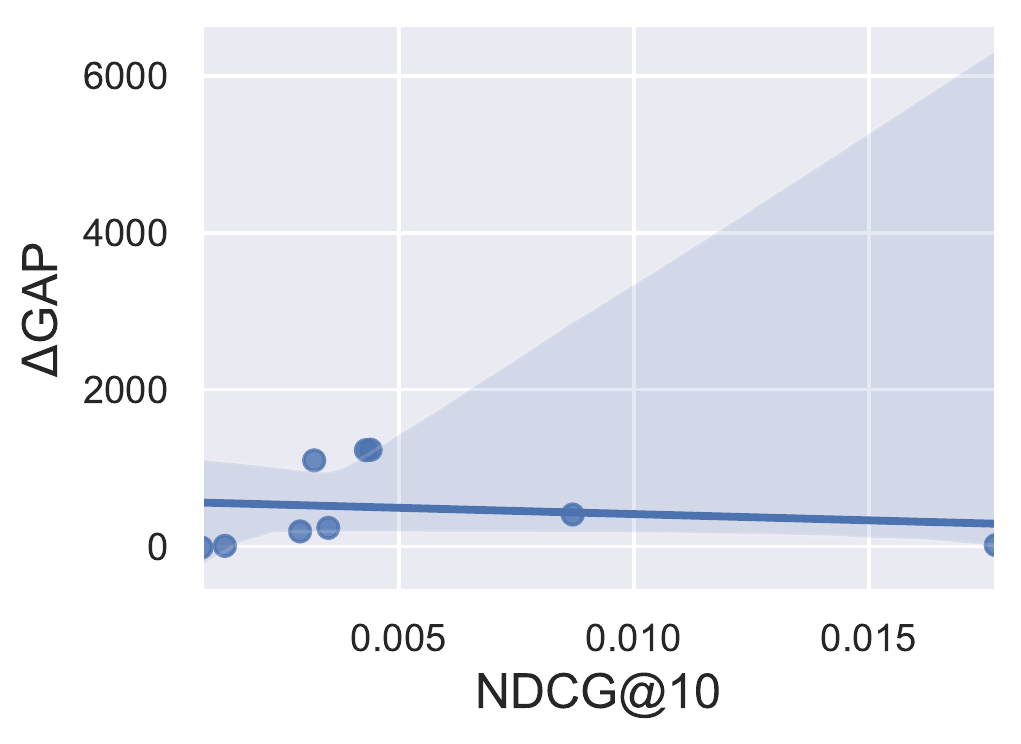}
            }
        }
    \subfloat[Diverse]
        {
            {
                \includegraphics[scale=0.37]{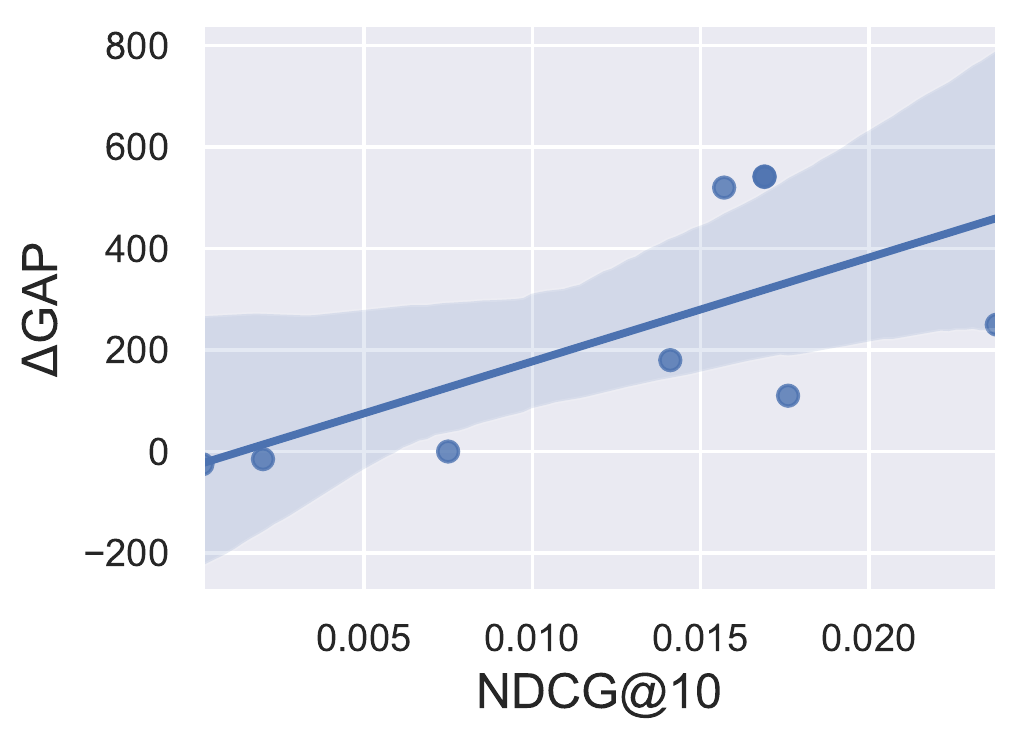} 
            }
        }
    \subfloat[Bestseller-focused]
        {
            {
                \includegraphics[scale=0.37]{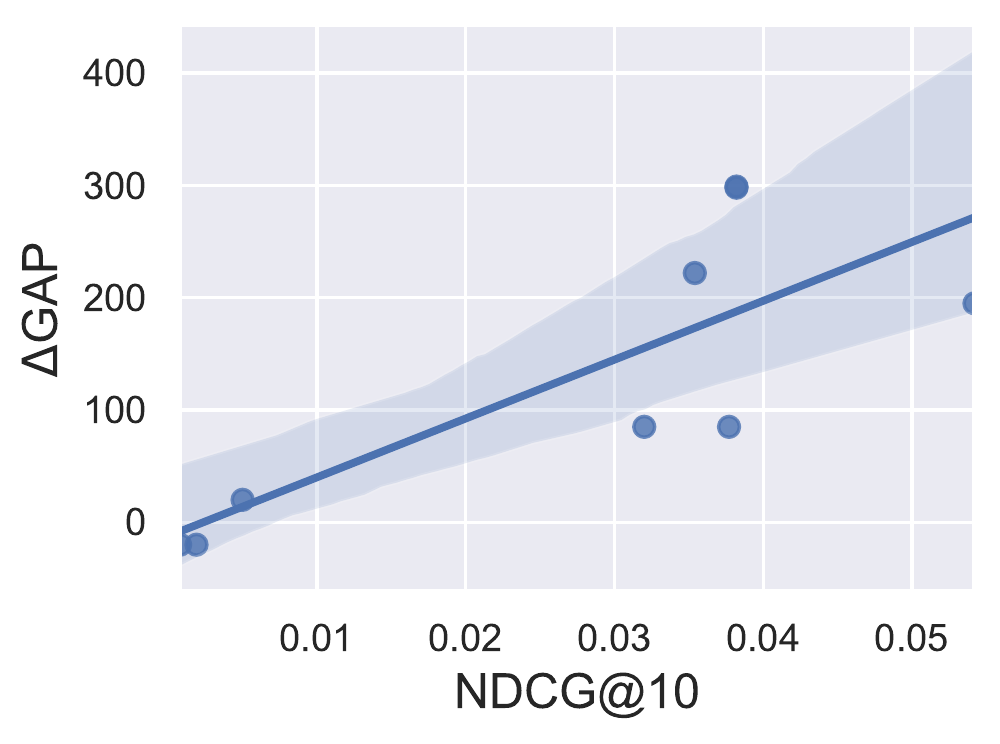} 
            }
        }
    \caption{The correlation between NDCG and $\Delta$GAP for the three user groups.}
    \label{fig:tradeoff}
\end{figure}

Then, to address \textbf{RQ2}, we analyze the results of MAE (the lower, the better), Precision, Recall, and NDCG (the higher, the better) of the recommendation algorithms on three users groups. Moreover, we determine the statistically significant differences using the two-tailed paired t-test at a $95\%$ confidence interval ($p < 0.05$).
As we see in Fig.~\ref{fig:groups_results}, the Niche group receives significantly worse recommendations than two other groups (\ie Diverse and Bestseller-focused), while the Bestseller-focused group gets the best performance. According to this result, algorithms are not capable of detecting the difference in taste between users, even though the average size of user profiles for Niche users is larger than that of the Bestseller-focused group (\ie more training data), further emphasizing the unfairness of popularity bias. Across the algorithms, we see that \wmf and \vaecf algorithms provide the highest accuracy in all user groups. Notably, the \wmf algorithm displayed the best performance in Niche group when considering both accuracy and fairness.
\subsection{Unfairness of Popularity Bias vs. Personalization}
The main objective of this part is to explore the potential correlation or trade-off between unfairness of popularity bias and personalization measured by NDCG, between groups of users who have differing preferences for popular items. To this purpose, Fig.~\ref{fig:tradeoff} shows the correlation plot between NDCG@10 and $\Delta$GAP for three user groups defined in Section \ref{user_profile_size_and_popularity_bias_in_book_data} where each point represents the performance of a state-of-the-art recommendation algorithm. Interestingly, we observe an uphill (positive) correlation between NDCG and $\Delta$GAP in the Bestseller-focused and Diverse user groups with the p-value and Pearson's coefficient of (0.01, 0.79) and (0.05, 0.66), respectively. In contrast, we find no meaningful correlation (\ie p-value of 0.69) between accuracy and unfairness of popularity bias among users with Niche tastes. These results indicate that algorithms with a high accuracy score fall short on popularity bias fairness from the perspective of users with Diverse and Bestseller-focused tastes, prompting further studies on how to incorporate users' taste and expectations in recommendation without sacrificing the overall accuracy.

\section{Discussion}
\label{sec:discussion}
In this section, we present a summary of the answers we found to the research questions in Section \ref{sec:intro}.
\begin{itemize}
    \item \textbf{Answer to RQ1}: In line with the previous study of Abdollahpouri et al.~\cite{abdollahpouri2019unfairness}, our experiments demonstrate that different users have a considerably different tendency towards popular items. Moreover, we discovered that around 83\% of users have read at least 20\% of unpopular books in their profile and expect to receive some of these items in recommendations. Our result further reveals that users with larger profile sizes who contribute most to the system have diverse tastes and interact with a substantial amount of unpopular items.
    \item \textbf{Answer to RQ2}: Our results for various state-of-the-art recommendation algorithms demonstrate that most algorithms are unfair to users who have a niche or diverse taste in books in terms of popularity, \ie these users receive recommendations that have lower accuracy and mainly consist of popular books. In addition, the study shows popularity bias negatively affects all user groups, even those focusing on bestsellers, but the magnitude of this effect varies greatly depending on the user group.
\end{itemize}

\section{Conclusion and Future Work}
\label{sec:conclusion}
In this paper, we reproduced the study of Abdollahpouri et al.~\cite{abdollahpouri2019unfairness} on the unfairness of popularity bias from the user's perspective in the Movie domain, which we have applied to the book domain. Similar to the original paper, we divided all users into three groups (\ie Niche, Diverse and Bestseller-focused) based on their level of interest in popular items. Our results on various state-of-the-art recommendation algorithms reveal that the most widely adopted algorithms fail to capture users' interest in unpopular items and recommend mostly popular items. Notably, the quality of recommendations received by users with a Diverse or Niche taste is significantly lower than that of users with Bestsellers taste, despite having a large profile size. Moreover, our experiments led to new observations and possible directions for future research. 
First, we noticed that algorithms could differ significantly in their ability to capture users' tastes based on the domain. For instance, the \nmf algorithm suffers from the unfairness of popularity bias in the music domain~\cite{kowald2020unfairness} while offering an entirely fair recommendation in \dataset dataset. A future research direction that would be interesting is identifying the underlying reason for the variance, in particular, which feature of the data (e.g., sparsity, average user interaction) plays the primary role in propagating the popularity bias.
Additionally, our results suggest that an underlying tradeoff exists between personalization and fairness of popularity bias in Diverse and Bestseller-focused groups, that is, algorithms with high personalization abilities tend to experience fairness issues. Thus, further research could be worthwhile into implementing a recommendation algorithm that can find the optimal tradeoff between personalization and the unfairness of popularity biases to enhance the system's overall effectiveness. Finally, it would be interesting to investigate popularity bias on other domains and algorithms such as session-based \cite{wang2021survey}, content-based \cite{lops2011content}, or reinforcement learning-based recommendation \cite{afsar2021reinforcement} methods, as well as incorporating further evaluation metrics such as novelty and coverage.


\partitle{Reproducibility.}
To enable reproducibility of the results, we provide our dataset, source codes with all used parameter settings, and more experimental results and analysis on our webpage: \url{https://rahmanidashti.github.io/FairBook/}
%
%
%
\bibliographystyle{splncs04}
\bibliography{mybib}

\end{document}